\documentclass{jpp}
\usepackage{url}
\usepackage{hyperref}
\usepackage{graphicx}
 \usepackage{epstopdf, epsfig}
\usepackage[caption = false]{subfig}
\usepackage[utf8]{inputenc}
\usepackage[T1]{fontenc}
\usepackage{amsmath}

\title{Experimental study of argon gas breakdown with symmetric and asymmetric electrode configurations}
\author{Hridya P,  Mangilal Choudhary }
\affiliation{Department of Physics and Astrophysics, University of Delhi, Delhi-110007, India}
\date{17-10-2025}
\begin{document}
\maketitle
\begin{abstract}
Paschen’s law relates the breakdown voltage ($V_B$) of a gas to the product of gas pressure ($p$) and inter-electrode distance ($d$), predicting a characteristic minimum voltage at a specific $pd$ value. In this study, the role of electrode configurations (symmetric and asymmetric) and inter-electrode spacing on the gas breakdown processes (or Paschen Law) is examined. Numerous sets of experiments are performed with both symmetric and asymmetric electrode configurations of different sizes to obtain Paschen curves at different inter-electrode distances. The experimentally obtained Paschen's curves for different electrode configurations are fitted using a proposed modified empirical relation for $V_B$, incorporating variable power-law dependencies and fitting parameters to better capture the observed deviations. Upon closer inspection, we observed that the breakdown voltage ($V_B$) and the corresponding $pd$ value ($pd_{min}$) are influenced by both electrode configurations and inter-electrode discharge gap. The variation in $V_B$ and $pd_{min}$ for different electrode configurations is explained by analyzing the electric field distributions between the electrodes (cathode and anode) for an applied voltage. 
\vskip 2mm
\end{abstract}
\maketitle

\section{Introduction}\label{s:1}

Glow discharge plasma is one of the most straightforward and effective methods for producing plasma in laboratory settings, characterized by distinct regions between the cathode and anode where the gas transitions from a dielectric to a conducting state once the applied voltage exceeds a critical threshold \citep{bogaerts1996mathematical}. It has a wide range of applications, notably in materials processing for surface modification, thin-film deposition, etching, and doping, as it provides a stable and controllable source of ions and radicals \citep{venugopalan1984analysis, surface_modification, akther2023insights, zhang2013effect, yang2007dc, aronsson1997glow, yari2022glow}. Also glow discharge plasmas are leveraged in catalytic processes for environmental and chemical synthesis applications \citep{van2025plasma}, plasma-enhanced chemical vapor deposition, cleaning and activation of surfaces \citep{aronsson1997glow}. Additionally, such plasma systems are widely utilized in analytical spectroscopy for elemental analysis \citep{bogaerts1999glow}, in lighting (fluorescent and neon lamps), manufacturing of semiconductors due to their ability to produce uniform plasma conditions, and in plasma catalysis reactors for efficient CO2 conversion and other catalytic reactions \citep{hammer2004plasma, dkebek2019low}. The versatility of glow discharge plasma in industrial, environmental, and biomedical fields makes it an important plasma source.\\
In the discharge chamber, free electrons, originating from field emission or cosmic rays, initiate the discharge process. Once an electric field is applied, these electrons accelerate along the field lines, gaining energy and undergoing repeated collisions with neutral gas molecules. Each collision may transfer enough energy to ionize the neutral species, releasing additional electrons. This cyclic process, governed by the mean free path ($\lambda$) and characterized by an average drift velocity, leads to sustained electron multiplication and the development of a self-sustaining discharge. Breakdown and discharge characteristics depend on properties such as gas type, gas pressure, and distance between electrodes \citep{pdcurvefirst1889}.\\
Paschen's law, formulated by the German physicist Louis F. Paschen in 1889, describes the relationship between the breakdown voltage ($V_B$) required to ionize a gas and the product of the gas pressure ($p$) and the inter-electrode distance ($d$), i.e., $V_B = f(pd)$\citep{lisovskiy2000low}. 
Townsend introduced three coefficients to explain the breakdown process and the breakdown criterion\citep{townsend1915electricity, posin1936townsend}.
The first Townsend coefficient ($\alpha$) is the probability per  unit length of path to ionize neutral gas by electron impact collision and is expressed as
\begin{equation}
    \alpha = \frac{1}{\lambda_e}\exp{\frac{-E_i}{\epsilon_e}}
    \label{alpha eqn}
\end{equation}
where $\lambda_e$ is the mean free path, 
\begin{equation}
    \lambda_e = \frac{k_BT}{\sigma_ip},
    \label{mean free path}
\end{equation}
$\sigma_i$ being the electron impact ionization cross section, $E_i$ is the gas ionization threshold, and $\epsilon_e$ is the energy gained by the electron between two consecutive collisions with neutral atoms, and is assumed to be ($\epsilon_e = eE\lambda_e$)\citep{talviste2021experimental}. For a uniform electric field, the electric field before breakdown is taken as $E=\frac{V}{d}$ where V is the applied voltage and d is the inter-electrode distance, neglecting the space charge effects. Including all these in Eq.(\ref{alpha eqn}) , $\alpha$ can be written as,
\begin{equation}
    \alpha = A p \exp({\frac{-Bp}{E}})
    \label{alpha}
\end{equation}
where $A=\frac{\sigma_i}{k_bT}$ and $B= \frac{\sigma_iE_i}{ek_BT}$ are gas-dependent constants\citep{burm2007calculation}.
The second ionization coefficient ($\beta$) is the ionization caused by positive ions during breakdown. As the energy of the positive ions is much less than the energy of electrons, in most cases they are considered to be zero, making this the least important among the three. The third Townsend coefficient or the effective secondary emission coefficient ($\gamma_{se}$) describes the electron emission due to the impact of particles bombarding on the cathode\citep{sen1962variation, donko2001apparent, auday1998experimental}. The particles can be ions, metastable atoms, energetic electrons or photons and fast neutrals. i.e., $\gamma_{se}$ can be written as:
\begin{equation}
    \gamma_{se} = \gamma_{ion} +\gamma_{meta } + \gamma_{electron} + \gamma_{neutrals } +\gamma_{photon } 
\end{equation}
The secondary electron emission can be due to multiple factors, but the ion impact is the dominant one.
So, combining the exponentially increasing electron number due to avalanche processes and the secondary electron emission processes.  If the ionization threshold energy of the gas is higher than the work function of the cathode, the $\gamma_{se}$ due to ion impact will be higher.
Combining the avalanche of exponentially growing electron number and the secondary electron emission processes from cathode, the total current density can be expressed as,
\begin{equation}
    J = \frac{J_0 \exp{(\alpha d)}}{1-\gamma_{se}(\exp{(\alpha d)}-1)}
    \label{current density}
\end{equation}
where $J_0$ is considered to be the initial current near the cathode surface, due to the cosmic rays.
For a self-sustaining discharge, the total current will be a nonzero value even if $J_0$ vanishes, i.e., the denominator of Eq.(\ref{current density}) should be zero, introducing the Townsend breakdown criterion:
\begin{equation}
    \gamma_{se} (\exp{(\alpha d)}-1) =1
    \label{breakdown condition}
\end{equation}
Here, $\gamma_{se}$ and $\alpha$ depend on external voltage.
Rearranging,
\begin{equation}
    \alpha d = \ln{(1+\frac{1}{\gamma_{se}})}
    \label{alpha d townsend condition}
\end{equation}
By comparing Eq.(\ref{alpha}) and Eq.(\ref{alpha d townsend condition}), the breakdown voltage can be expressed as
\begin{equation}
    V_B = \frac{B pd}{\ln{(Apd)}-\ln{(\ln{(1+\frac{1}{\gamma_{se}})})}}
    \label{paschen law OG}
\end{equation}
The above equation is the mathematical form of the Paschen law \citep{Fu_2020, loveless2017universal} which can be also be written as:
\begin{equation} \label{eq:9}
    V_B = \frac{Bpd}{\ln{(pd)}+C}
\end{equation}
where $C= \ln{\frac{A}{\ln{(1+\frac{1}{\gamma}})}}$ \citep{husain1982analysis}. In literature, the values of the coefficients for argon gas are reported as $A \approx$ 10.20 mbar$^{-1}$cm$^{-1}$ and $B\sim$ 176 V mbar$^{-1}$cm$^{-1}$\citep{burm2007calculation, cobine1958gaseous}.\\
\begin{figure}
    \centering
    \includegraphics[width=0.8\linewidth]{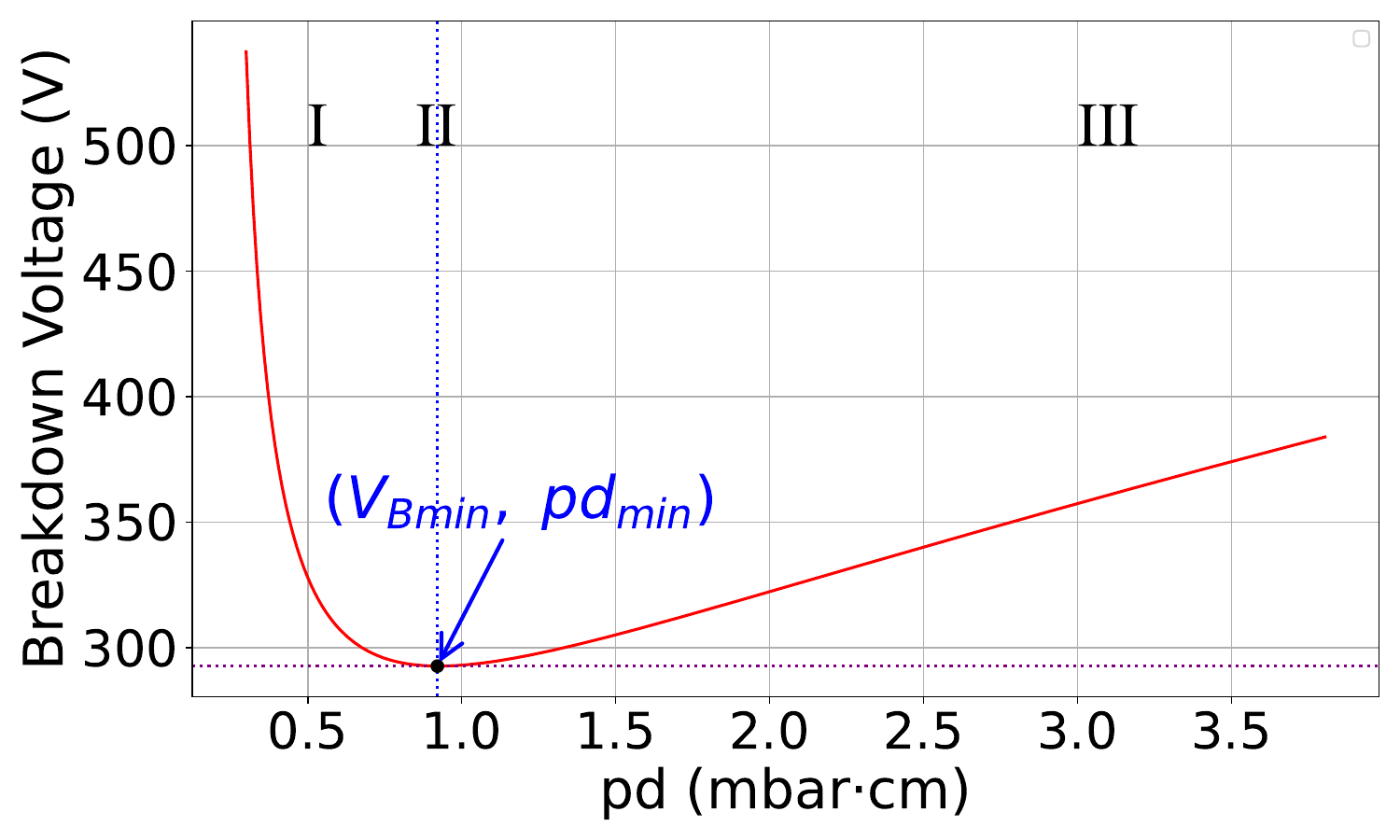}
    \caption{A general Paschen's curve for gas breakdown}
    \label{fig:Fig1}
\end{figure}
 The graphical representation of eq.~\ref{eq:9} is depicted in Fig.\ref{fig:Fig1}. This V-shaped curve is called as Paschen's curve and has three distinct regions (see Fig.\ref{fig:Fig1}). At lower $pd$ values (Region-I), the gas pressure in the chamber or the inter-electrode distance, or both, will be low, causing a longer electron collisional mean free path. Thus, the breakdown of gas occurs at a higher discharge voltage. As $pd$ increases, the mean free path decreases and reaches a minimum point, at which the breakdown voltage is also minimum (Region-II). Further increasing $pd$ leads to frequent collisions, making the electrons lose energy rapidly. Therefore, a higher voltage is required to accelerate electrons sufficiently to achieve ionization, despite frequent energy losses leading to an increase in the breakdown voltage (Region-III).\\
 
Townsend mentions that Paschen's law is just a special case of the uniform electric field, of a more general similarity theorem, which can be applied also to non-uniform E-fields, if the breakdown depends on ionization due to collision between electrons and neutral atoms \citep{townsend1915electricity}. The similarity theorem is explained in detail by Von Engel\citep{engel1965ionized, fu2016similarity}. Studies back in 1928 have reported that the breakdown voltage actually depends independently on pressure and interelectrode distance, i.e., $V_B =f(p,d)$ rather than $V_B =f(pd)$ \citep{intersection_pd_2016, penning19341007, Townsend01111928, boyle1955departure}. Some recent studies suggest that the size of the electrodes also affects the breakdown process, and the breakdown voltage can be expressed as $V_B = f(pd,\frac{d}{r})$, where $r$ is the radius of the electrode. This is also considered as a special case of the above-mentioned similarity theorem, of a non-uniform field between parallel-plate electrodes\citep{osmokrovic2006validity, kartalovic2023validity, mezei1998similarity, dekic2010conditions}.\\
Ledernez \textit{et al.} (2012) refined Paschen’s law using low-pressure argon experiments and Townsend's theoretical analysis, demonstrating that the breakdown voltage depends not only on the product \(pd\) but also independently on the electrode gap \(d\) \citep{interelectrodepd2012, mathew2019experimental}. They introduced an effective secondary emission coefficient $\gamma_{eff}$ to replace the classical \(\gamma_{se}\), accounting for various emission mechanisms such as ion, metastable, photon, electron, and neutral particle induced emissions.
Palomares \textit{et al.} (2022) experimentally studied ethanol (CH\(_3\)CH\(_2\)OH), methanol (CH\(_3\)OH), and their mixtures, observing that Paschen curve parameters \(A\), \(B\), and the secondary emission coefficient \(\gamma\) vary with electrode geometry, specifically on the electrode radius ratio \(\frac{d}{r}\) \citep{ethanolpaschen}. Notably, the 50\% ethanol–methanol mixture yielded the lowest breakdown voltage, attributed to combined molecular ionization potentials and enhanced secondary emission.
Lyu \textit{et al.} (2020) modified the Paschen curve for large-area, short glow discharges by generalizing the breakdown voltage formula from the classical form Eq.(\ref{paschen law OG}) to
\begin{equation} 
    V_B = \frac{B' (pd)^\chi}{\ln \left[ A' (pd)^\delta \right] - \ln \left[ \ln\left(1 + \frac{1}{\gamma} \right) \right]},
    \label{Modified_paschen_lyu}
\end{equation}
where \(\chi \leq 1\), \(\delta \leq 1\), \(\gamma = \frac{\gamma_{se}}{\zeta}\),($\zeta$ accounts for the effects of surface structure and cleanliness) and the normalized coefficients are defined as
\[A' = A \frac{p_0 d_0}{(p_0 d_0)^\delta}, \quad B' = B \frac{p_0 d_0}{(p_0 d_0)^\chi}, \quad p_0 d_0 = 1 \text{ Pa·m}.\]
This formulation accounts for non-uniform fields and electrode geometry effects \citep{lyu2020paschen}. \\
Additional verification and refinement of modified Paschen laws have been reported by Prijil \textit{et al.} (2019), showing breakdown voltage variation with electrode geometry beyond the \(pd\) product \citep{mathew2019experimental}. Furthermore, the effects of temperature, electrode material, and field emission at microscale distances have been studied extensively \citep{fu2016similarity, go2010mathematical, radwan2023modification, fu2017effect, galli2019paschen, sili2011temperature, abd2014experimental, torres2012paschen, brayfield2019impact, tirumala2010analytical, 4073800}.
These results collectively suggest a generalized breakdown voltage relation:
\begin{equation}
    V_B = f\left(pd, d, \frac{d}{r}, T, \text{geometry}\right),
\end{equation}
which reduces to classical Paschen’s law under uniform electric fields and typical electrode configurations.
\\\\
 It is a fact that extensive studies have been conducted to understand the electric breakdown of different gases under different experimental conditions. However, there are still many open questions regarding the validity of Paschen’s law in cases of non-uniform electric fields with different-sized electrode configurations. Is there a threshold inter-electrode distance up to which  Paschen’s law is valid in the case of a symmetric or asymmetric electrode configuration? How does the minimum breakdown voltage vary with increasing inter-electrode spacing in symmetric and asymmetric electrode configurations? How does the $pd_{min}$ correspond to the breakdown voltage vary with increasing inter-electrode spacing in symmetric and asymmetric electrode configurations? How much additional voltage is required to initiate the breakdown in the case of different geometric-sized electrode configurations? To address these questions, numerous experiments with symmetric and asymmetric electrode configurations (cathode and anode) are performed in a newly built direct current (DC) discharge plasma device. The results reveal that the existing Paschen law does not adequately describe the breakdown characteristics for these electrode geometries at different inter-electrode spacings. Furthermore, the commonly proposed modifications to Paschen's law, which incorporate additional parameters or adjustment factors, also fail to accurately explain the observed experimental behaviour. Consequently, a more general form of Paschen's law was proposed to account for the effects of electrode configurations and inter-electrode spacings. We have also examined the variation of $pd_{min}$ and the breakdown voltage as a function of the inter-electrode distance, highlighting the influence of electrode asymmetry and size on the discharge behaviour.\\
A detailed description of the experimental set-up and pd curve measurements is given in Sec.\ref{sec:sec2}. The experimentally observed pd curves, variation in breakdown voltage and shifting of $pd_{min}$ for symmetric and asymmetric electrode configurations are presented in Sec.\ref{sec:sec3}. The experimentally observed results are explained in Sec.\ref{sec:sec4}. A summary of the work and concluding remarks are provided in Sec.~\ref{sec:sec5}.
\section{Experimental device and measurement} \label{sec:sec2}
\begin{figure}
    \centering
    \includegraphics[width=0.90\linewidth]{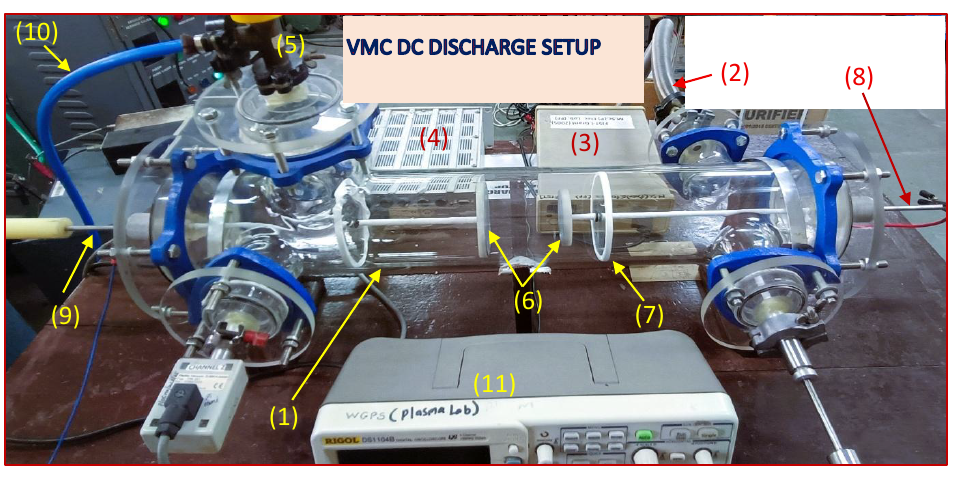}
    \caption{A DC Discharge plasma setup: (1) Borosilicate glass tube, (2) rotary pump, (3) digital ammeter, (4) pressure gauge, (5) needle valve, (6) electrodes, (7) supporting insulating disk, (8) and (9) SS-304 rods, (10) argon gas pipe connected to cylinder, and (11) oscilloscope.}
    \label{fig:Fig2}
\end{figure}

\begin{figure}
    \centering
    \includegraphics[width=0.90\linewidth]{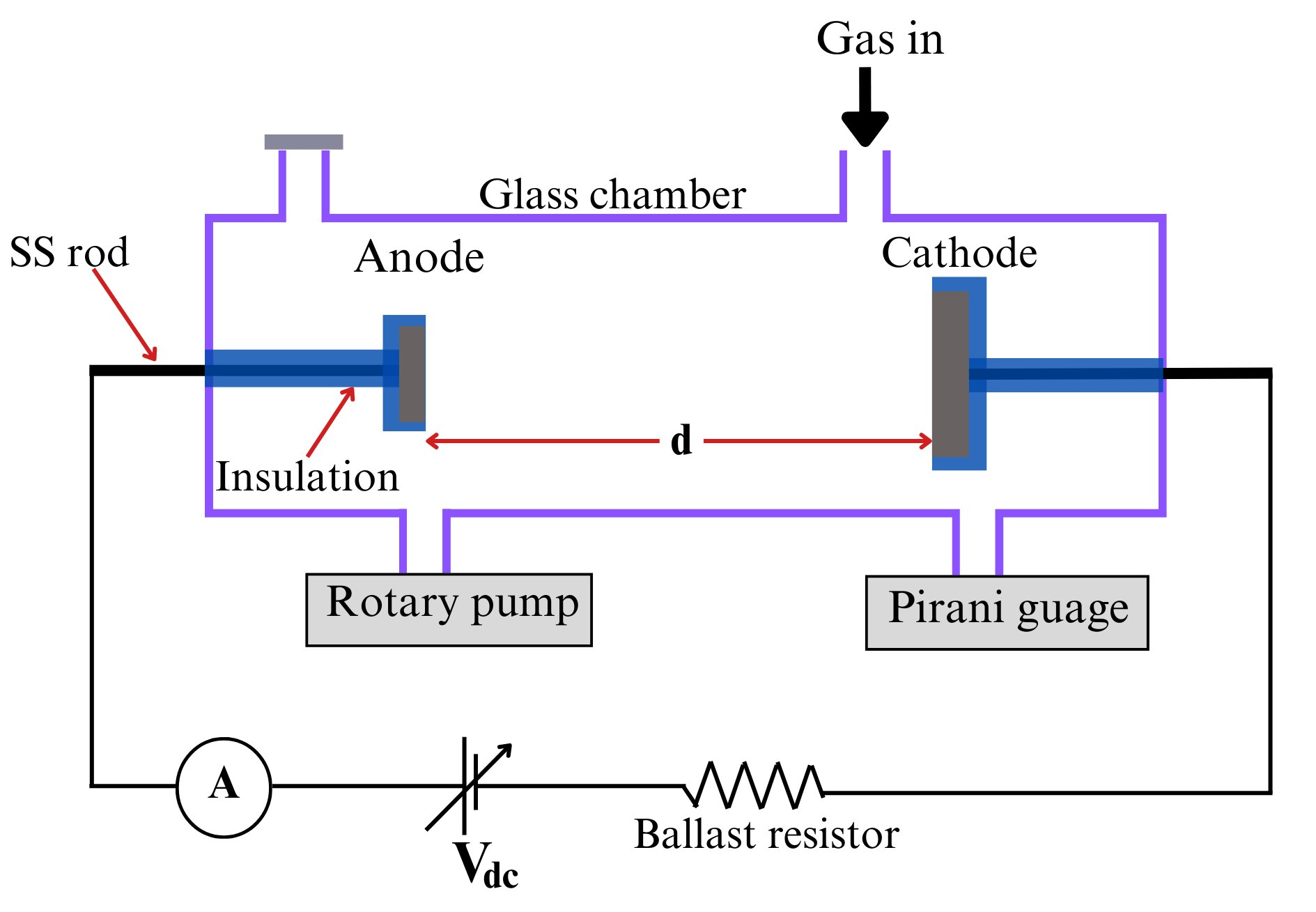}
    \caption{Schematic diagram of experimental setup}
    \label{fig:Fig3}
\end{figure}

The experimental studies are conducted in a newly built in-house VMC DC (direct current) discharge plasma device at the Plasma Physics Laboratory. A photograph of the VMC DC discharge plasma device is depicted in Fig.\ref{fig:Fig2}. The device consists of a cylindrical borosilicate glass tube, 70 cm in length and 10 cm in radius, with 5 radial and 2 axial ports. The end ports of the tube are used to fix the cathode and anode, while the radial ports are utilized to evacuate the glass chamber through a rotary pump, measure the gas pressure using a Pirani gauge, and feed gas through a needle valve. The position of the cathode and the anode can be changed by moving the stainless steel 304 rods through the end-port flanges. A schematic diagram of the experimental setup is shown in Fig.\ref{fig:Fig3}. The cathode and anode of desired sizes are made from an aluminium sheet of 5 mm thickness. The rear surfaces of the electrodes are covered with heat-resistant insulating material, polytetrafluoroethylene, to prevent additional areas of the cathode and anode from being exposed. Two additional insulating (Acrylic) disks through SS rods are fixed inside the chamber to keep the electrodes parallel to each other at a larger distance. A DC power supply (1000 V, 1 A) is used to provide the voltage necessary to initiate the breakdown of argon gas at a given gas pressure (p) and inter-electrode distance (d). The glass tube is evacuated to the lowest base pressure ($\sim 10^{-3}$ mbar) and then injected argon gas to conduct experiments.  
\\\\
To measure the breakdown voltage at given values of $p$ and $d$, an ammeter is connected in series with the discharge path. A sudden jump in discharge current indicates the gas breakdown point, and the corresponding voltage is recorded as the gas breakdown voltage. The breakdown voltage is measured by keeping the inter-electrode spacing ($d$) constant and varying the gas pressure. The same exercise is repeated after changing the inter-electrode distance for a given cathode-anode size configuration. In this way, a Paschen curve ($pd$ curve) is constructed for a given electrode configuration. The same exercise is repeated three times on different days to confirm the consistency in observed $pd$ curves for a given electrode configuration. For another set of experiments with different electrode configurations, electrodes (cathode and anode) are replaced by opening the side ports of the device. It should be noted that we always used clean and smooth surface electrodes for all experiments and avoided possible conducting paths by using insulation. The conducting path is limited to the area between the front of the cathode and the anode.              

\section{Experimental Results} \label{sec:sec3}
For a detailed study of argon gas breakdown with symmetric and asymmetric electrodes of different sizes in an insulating glass tube at various pressures and inter-electrode spacings, numerous experiments are conducted. We define a "symmetric electrode configuration" as one in which the electrodes are similarly sized, and an "asymmetric electrode configuration" as one in which the anode is smaller in size than the cathode.\\
Experimental data obtained from various electrode configurations were initially fitted to the classical Paschen law (eq.~\ref{eq:9}) using coefficient values reported in literature. As anticipated, substantial deviations were observed, particularly at both small ($\sim$ 1 cm or 2 cm) and large ($\sim$ 30 cm) gap distances. So we moved on to apply modified versions of Paschen's law (eq.~\ref{Modified_paschen_lyu}), but this also failed to produce a better fit for the measured data. Through a series of empirical trials, a new modified relation was developed that offered an accurate fit for all tested configurations:
\begin{equation}
    V_B = \frac{B^*(pd)^k}{ln(pd)^m +C^*}
    \label{paschen new}
\end{equation}
where $B^*$ is a corrected form of the gas-dependent coefficient (originally denoted as $B$), having a value observed to vary across the range for different $d$ values. The parameter $C^*$ encompasses contributions from both the secondary electron emission coefficient and constant $A$, with values ranging from 1 to 15 over the measured $pd$ regime. Notably, the exponents $k$ and $m$ deviate from unity, indicating a dependence on $d$, which highlights the departure from the original Paschen equation and the presence of nonlinear effects. This generalisation accommodates the observed empirical deviations and provides a more accurate fit to the measured breakdown voltages across all electrode configurations examined.\\
The parameter $C^*$ in Eq.(\ref{paschen new}) is primarily responsible for the vertical offset of the entire Paschen curve, most noticeably affecting the left branch. As $C^*$ increases, the slopes of both the left and right branches decrease, and the minimum breakdown voltage shifts downward. Decreasing $k$ below unity selectively reduces the slope of the left branch, with little influence on other parts of the curve. Conversely, lowering $m$ from unity increases the left branch’s slope and also reduces both $pd_{\min}$ and the corresponding breakdown voltage. The parameter $B^*$ acts differently: increasing $B^*$ elevates $V_B$ while keeping $pd_{\min}$ fixed, thus shifting the minimum breakdown voltage upward without altering the location of the minimum on the Paschen curve.
\\
Experimentally observed Paschen breakdown curves ($pd$ curves) for symmetric and asymmetric electrode configurations, validation of proposed modified Paschen's Law (eq.~\ref{paschen new}), variation of $V_{B_{\min}}$ ($V_B$ corresponding to minimum $pd$) and $pd_{min}$ with inter-electrode gap ($d$) for different electrode configurations are discussed in the respective subsections. 
\subsection{Experiments with 80 mm diameter Cathode}
\begin{figure*}
 \centering
\subfloat{{\includegraphics[scale=0.3200050]{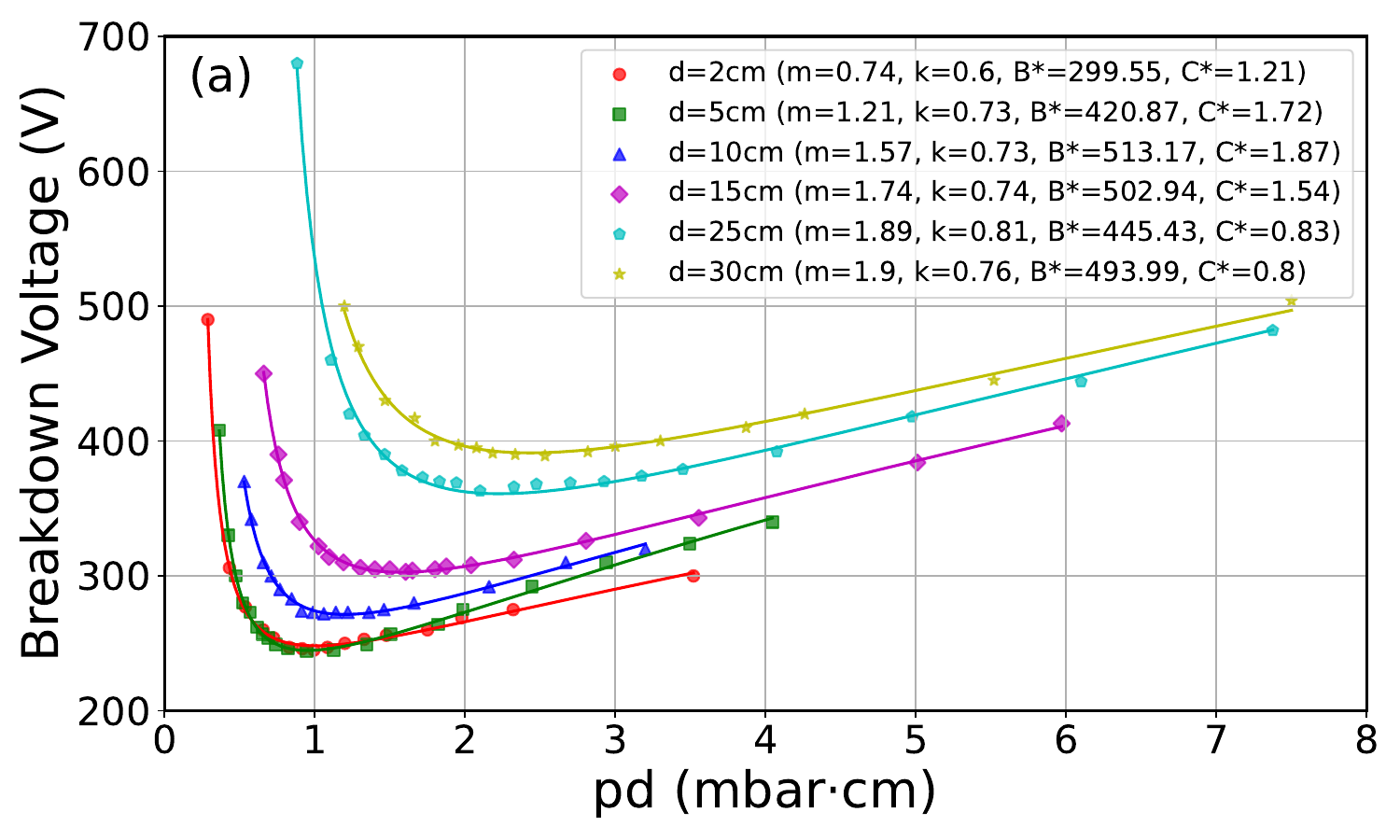}}}%
 \quad
 \subfloat{{\includegraphics[scale=0.320050]{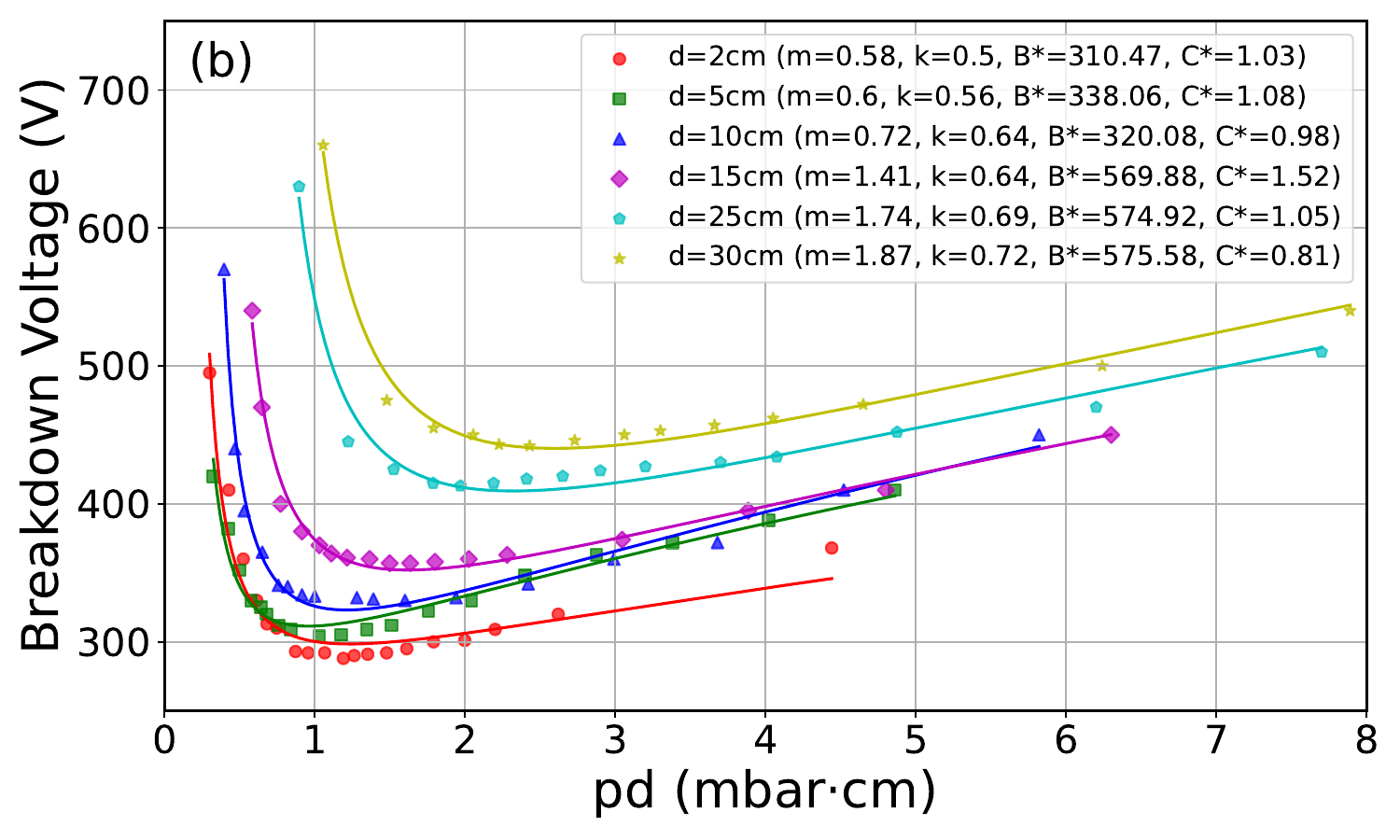}}}
\caption{\label{fig:Fig4} Fitted Paschen curves for (a) symmetric (80 mm cathode and 80 mm anode diameter) configuration and (b) asymmetric (80 mm cathode and 20 mm anode diameter) configuration at different inter-electrode distances.}
\end{figure*}
\begin{figure*} 
 \centering
\subfloat{{\includegraphics[scale=0.4000050]{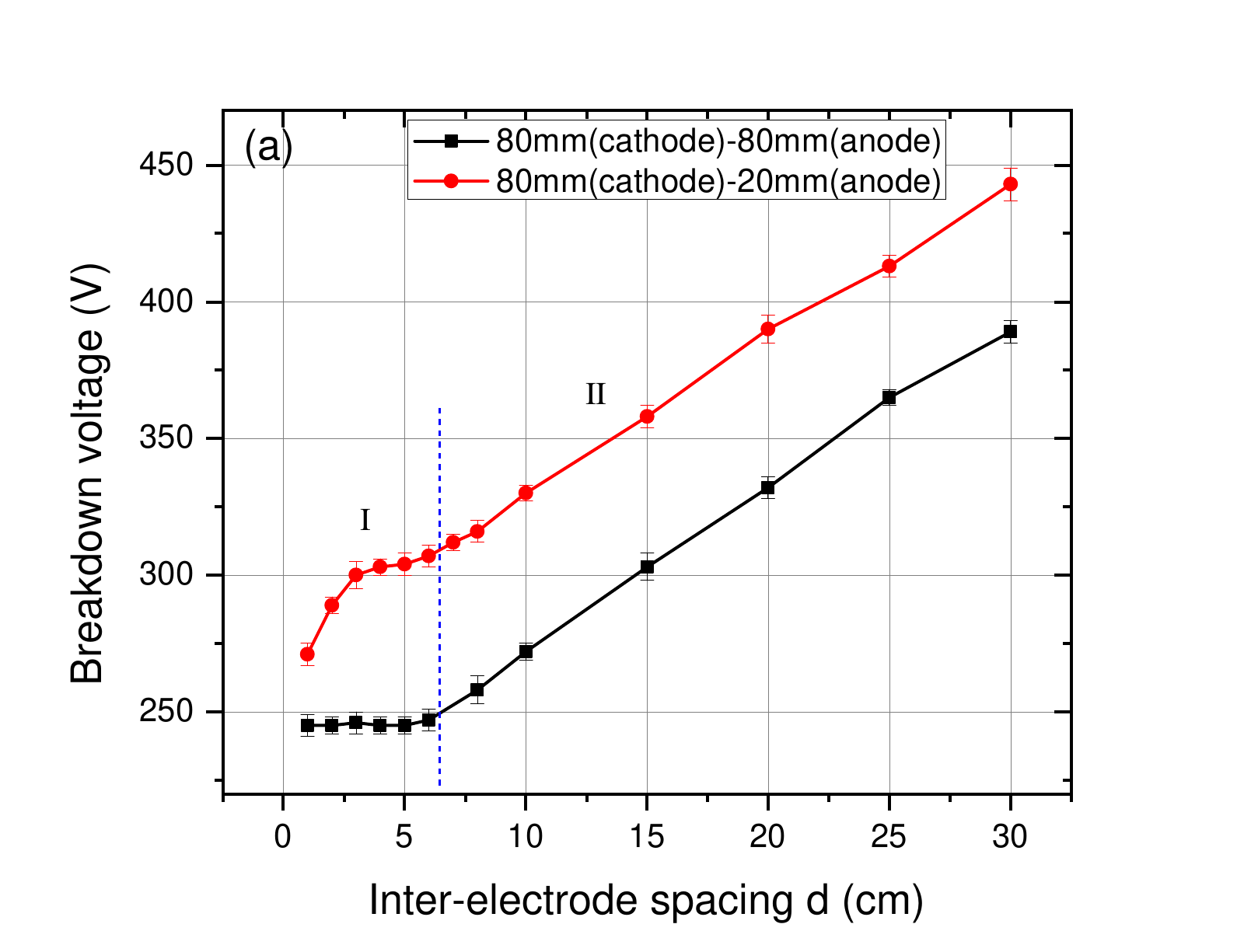}}}%
 \quad
 \subfloat{{\includegraphics[scale=0.400050]{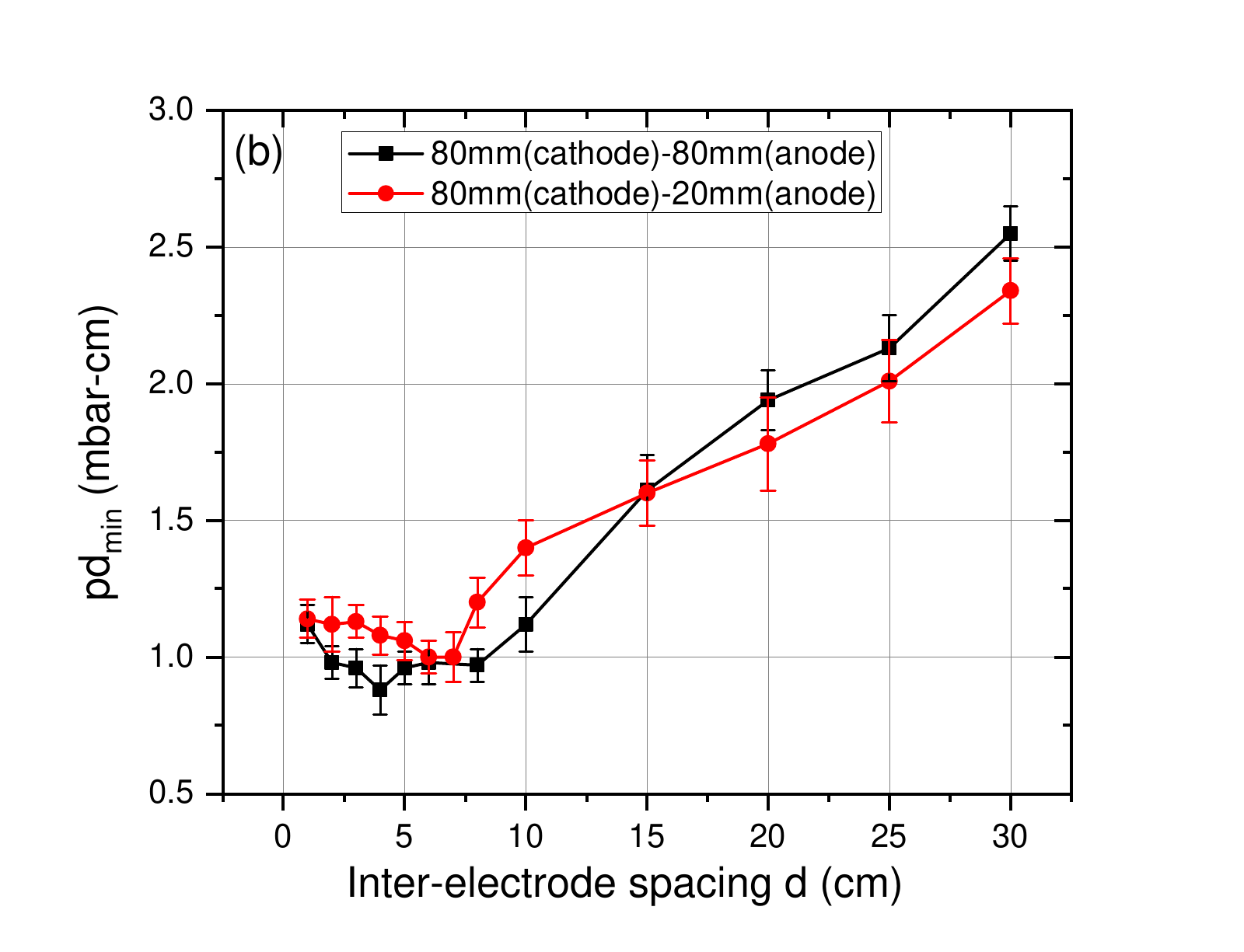}}}
\caption{\label{fig:Fig5}(a) Variation of $V_{B}$, corresponding to minimum value of $pd$ with inter-electrode spacing for symmetric (80 mm cathode and 80 mm anode diameter) and asymmetric electrode (80 mm cathode and 20 mm anode diameter). (b) variation of $pd_{min}$ with inter-electrode spacing for symmetrical and asymmetric electrode configurations.}
\end{figure*}

The sets of experiments are performed with a symmetric electrode configuration (cathode and anode of 80 mm diameter) and an asymmetric electrode configuration (cathode of 80 mm and anode of 20 mm diameter) to explore the role of both these configurations on gas breakdown. Fig.\ref{fig:Fig4}(a)  and Fig.\ref{fig:Fig4}(b) represent the variation of $V_B$ versus $pd$ for the symmetric and asymmetric electrode configurations, respectively. Across different electrode gaps, the proposed modified Paschen relation (Eq.~\ref{paschen new}) provides a better fit to the measured pd curves for all discharge gaps than classical formulations. Both configurations showed similar overall Paschen curve trends, but the Paschen minimum ($V_{B_{\min}}$, $pd_{min}$) is observed to shift in every curve.\\
For closer inspection, the $V_B$ corresponding to $pd_{min}$ at different inter-electrode distances are extracted from the obtained $pd$ curves for both electrode configurations. The variation of $V_B$ corresponding to $pd_{min}$ against $d$ for both configurations is displayed in Fig.\ref{fig:Fig5}(a). It is clear from Fig.\ref{fig:Fig5}(a) that $V_B$ remains nearly constant up to $d <$ 6 cm and then starts to increase approximately linearly up to $d =$ 30 cm. The argon gas breakdown occurs at a lower applied DC voltage ($\sim$250 V) in the symmetric configuration, where both electrodes are close to each other ($d < 80$ mm). However, $\sim$140 more volts are required to initiate the argon breakdown if both electrodes are placed 30 cm apart from each other, showing the influence of the inter-electrode gap on the breakdown process. \\  
However, the scenario is slightly different in the case of an asymmetric electrode configuration. We observe a sudden linear increase in $V_B$ up to $d<$ 4 cm, then it varies slowly between 4 to 6 cm and after that it increases linearly with increasing the electrodes gap up to $d$ = 30 cm. In the asymmetric configuration, gas breakdown happens at a higher DC voltage at any inter-electrode separation. The additional 25 to 50 volts are required to trigger the breakdown when electrode spacing is changed from $d =$ 1 cm to 30 cm [see Fig.\ref{fig:Fig5}(a)]. \\\\\
Fig.\ref{fig:Fig5}(b) deals with the variation of $pd_{min}$ against $d$ for both electrode configurations. The plots in Fig.\ref{fig:Fig5}(b) show that the value of $pd_{min}$ is slightly higher when electrodes are very close ($d <$ 2 cm) to each other. The $pd_{min}$ shifts towards a lower value and then again attains the same value as the spacing ($d <$ 10 cm) between cathode and anode increases. A nearly linear increase in $pd_{min}$ is observed beyond $d >$ 10 cm for both electrode configurations. However, there is no significant difference in $pd_{min}$ variation with $d$ for symmetric and asymmetric electrode configurations. Such a variation in $pd_{min}$ is reported for the first time in DC gas breakdown. For further verification, experiments were conducted with 40 mm and 20 mm cathode configurations.
\subsection{Experiments with 40 mm diameter Cathode}
\begin{figure*} 
 \centering
\subfloat{{\includegraphics[scale=0.4200050]{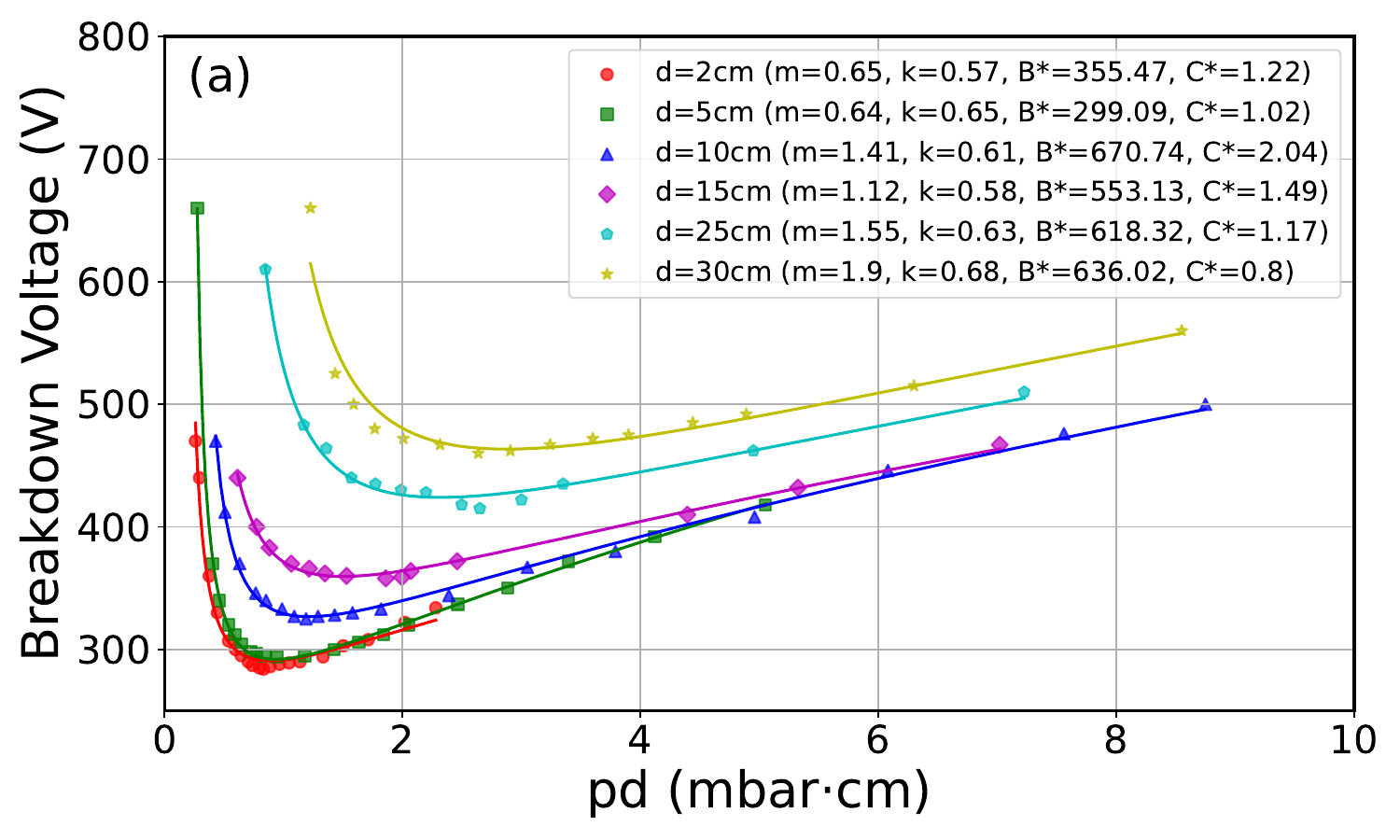}}}%
 \quad
 \subfloat{{\includegraphics[scale=0.420050]{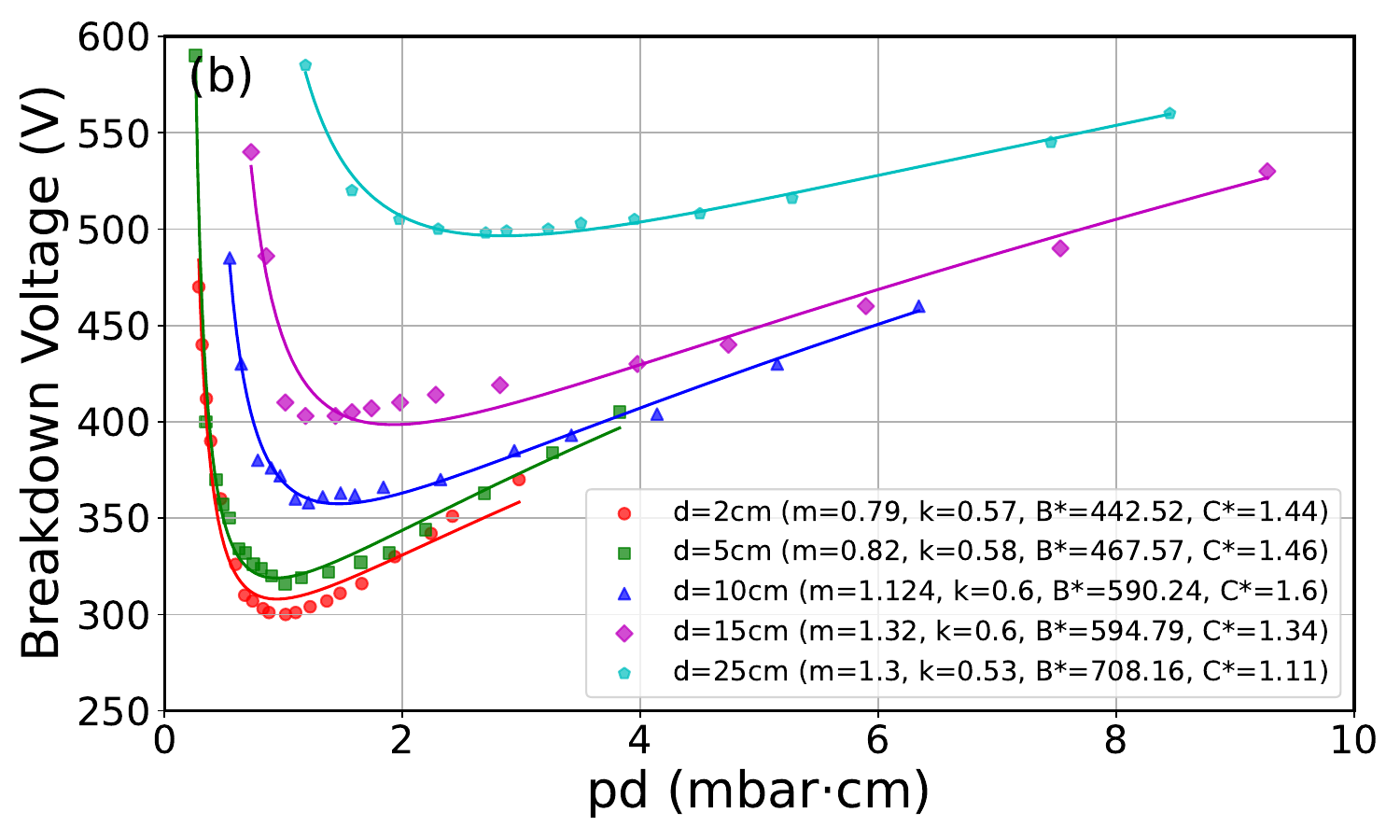}}}
\caption{\label{fig:Fig6} Fitted Paschen curves for (a) symmetric (40 mm cathode and 40 mm anode diameter) configuration and (b) asymmetric (40 mm cathode and 20 mm anode diameter) configuration at different inter-electrode distances.}
\end{figure*}

The second set of experiments utilized a cathode of 40 mm diameter paired with a 40 mm and 20 mm diameter anode for comparison of symmetric and asymmetric electrode configurations. The breakdown voltage against $pd$ values for the symmetric (40 mm - 40 mm)  and asymmetric (40 mm - 20 mm) electrode configurations is plotted in Fig.\ref{fig:Fig6}(a) and Fig.\ref{fig:Fig6}(b), respectively. All data are analyzed using the modified Paschen breakdown law, and the extracted curve parameters show modest deviations from those for the 80 mm cathode experiments, owing to the role of electrode size in argon gas breakdown.

\begin{figure*} 
 \centering
\subfloat{{\includegraphics[scale=0.4200050]{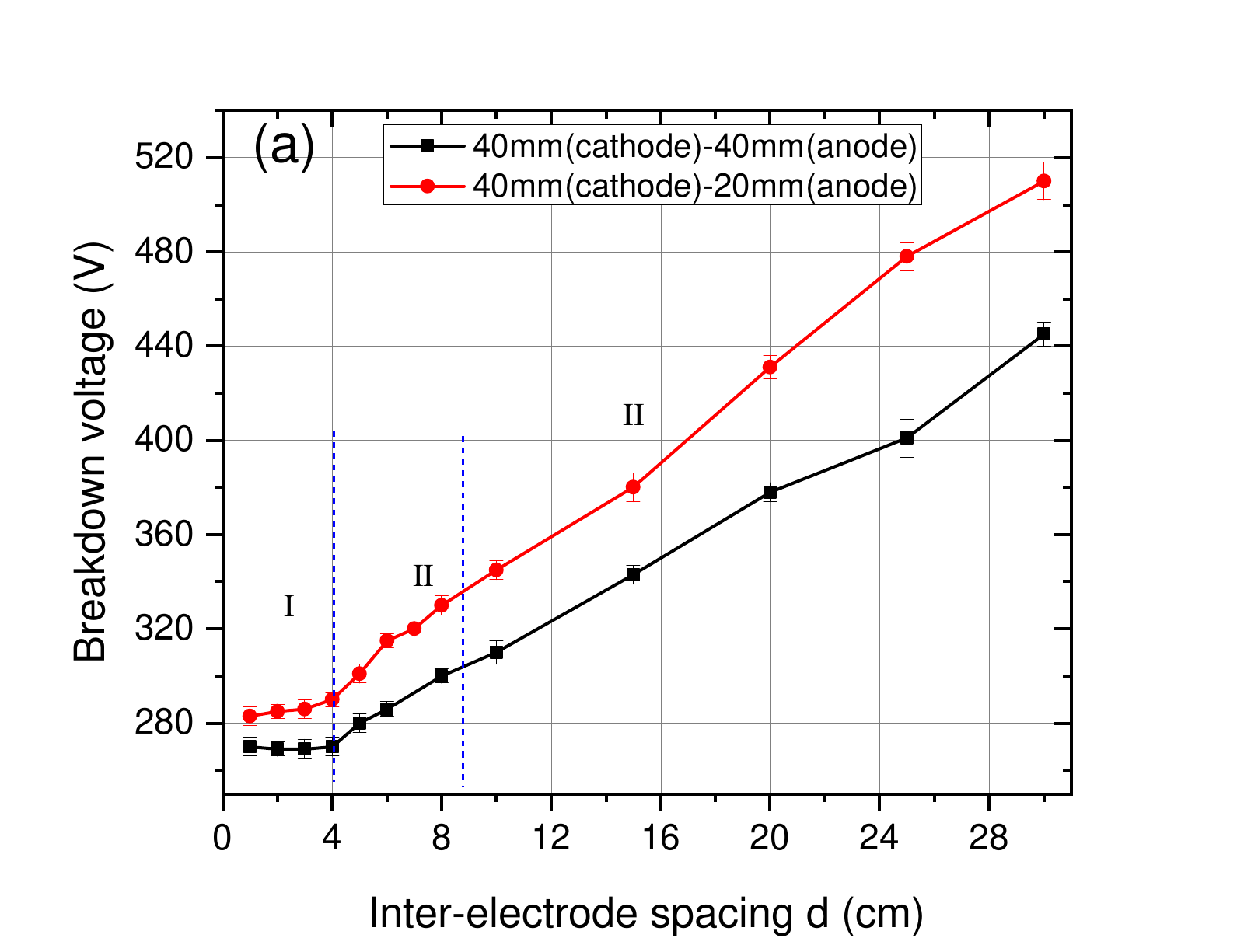}}}%
 \quad
 \subfloat{{\includegraphics[scale=0.420050]{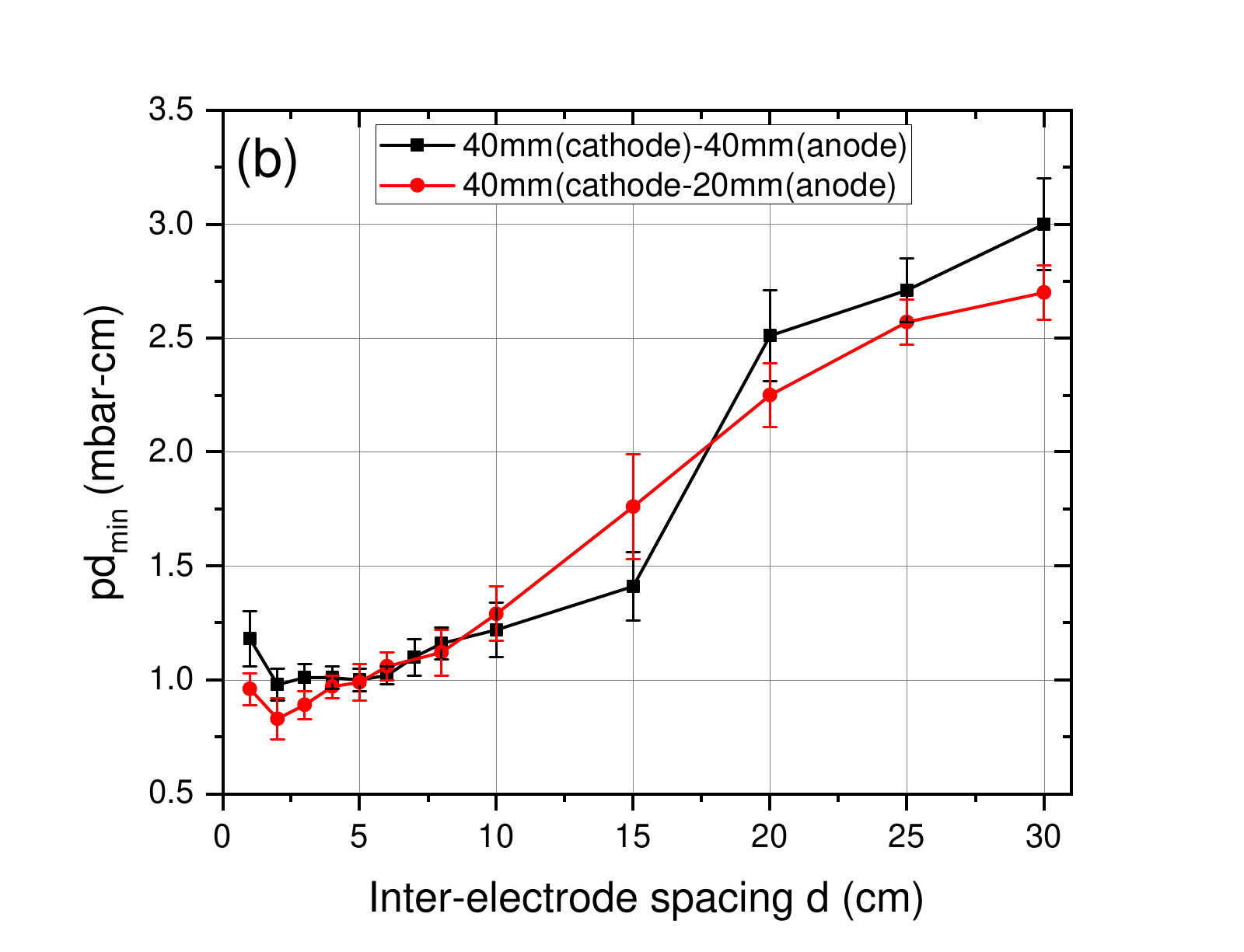}}}
\caption{\label{fig:Fig7}(a) Variation of $V_{B}$, corresponding to minimum value of $pd$ with inter-electrode spacing for symmetric (40 mm cathode and 40 mm anode diameter) and asymmetric electrode (40 mm cathode and 20 mm anode diameter). (b) variation of $pd_{min}$ with inter-electrode spacing for symmetrical and asymmetric electrode configurations.}
\end{figure*}
The $pd$ curves for symmetric and asymmetric electrode configurations are used to determine the minimum $V_{B}$ and $pd_{min}$ at different inter-electrode distances. For the symmetric case, $V_B$ remains nearly constant up to $d \leq$ 4 cm, and then it increases linearly with further increase in the inter-electrode spacing. However, the rate of increase of $V_{B}$ is observed to be slightly higher for $d<$ 10 cm compared to $d>$ 10 cm. Argon gas breakdown occurs at a lower applied DC voltage ($\sim$265 V) in the symmetric configuration when both electrodes are close to each other ($d <$ 4 cm). However, $\sim$170 additional voltage is required to initiate the breakdown of argon if both electrodes are placed 30 cm apart from each other.\\
In the asymmetric electrode configuration, we see a nearly 10 V difference in $V_{B}$ if the electrode spacing increases from 1 to 4 cm [see Fig.\ref{fig:Fig7}(a)]. Further increase in the inter-electrode spacing, $V_B$ corresponding to $pd$ minimum, increases linearly with slightly different rates between $d=$ 4 cm to 30 cm. The gas breakdown occurs at $\sim$280 V when both electrodes are very close to each other ($d$ = 1 cm). But we need a $\sim$500 V potential difference between cathode and anode when the inter-electrode spacing is about 30 cm. In this case, the breakdown voltage difference of $\sim$220 V is observed if the inter-electrode distance is increased from 1 cm to 30 cm. After comparing the minimum value of $V_{B}$ in Fig.\ref{fig:Fig7}(a) for both electrode configurations, it is concluded that gas breakdown occurs at a lower discharge voltage when both electrodes are of the same dimensions (40 mm in diameter). However, the value of $V_B$ does not remain the same for all inter-electrode spacings, contradicting the classical Paschen law.\\\\  
In Fig.\ref{fig:Fig7}(b), $pd_{min}$ is plotted against $d$ for both electrode configurations. We observe a shift in $pd_{min}$ from a slightly larger value to a minimum value as the electrode spacing increases from 1 to 2 cm. After that ($d>$ 2 cm), it increases with increasing $d$ and again attains the same value at $d\sim$ 10 cm. The value of $pd_{min}$ increases with increasing the inter-electrode distance from 10 cm to 30 cm for both electrode configurations. However, we do not record a significant difference in $pd_{min}$ variation with $d$ for symmetric and asymmetric electrode configurations.  The $pd_{min}$ follows a nearly similar trend, with a slight difference in its value at a given $d$ for the symmetric and asymmetric electrode configurations. 
\subsection{Experiments with 20 mm diameter Cathode}

\begin{figure*} 
 \centering
\subfloat{{\includegraphics[scale=0.4200050]{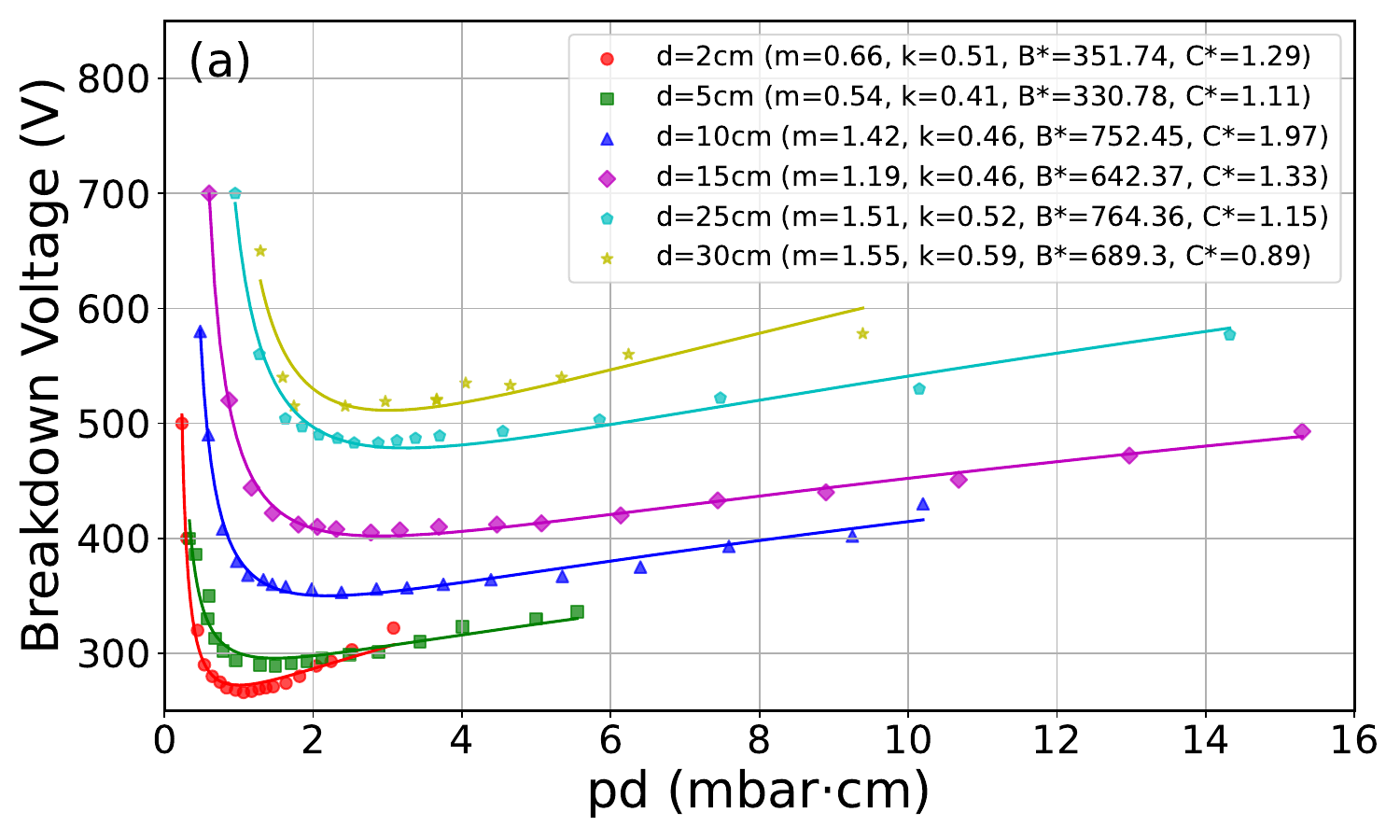}}}%
 \quad
 \subfloat{{\includegraphics[scale=0.4200050]{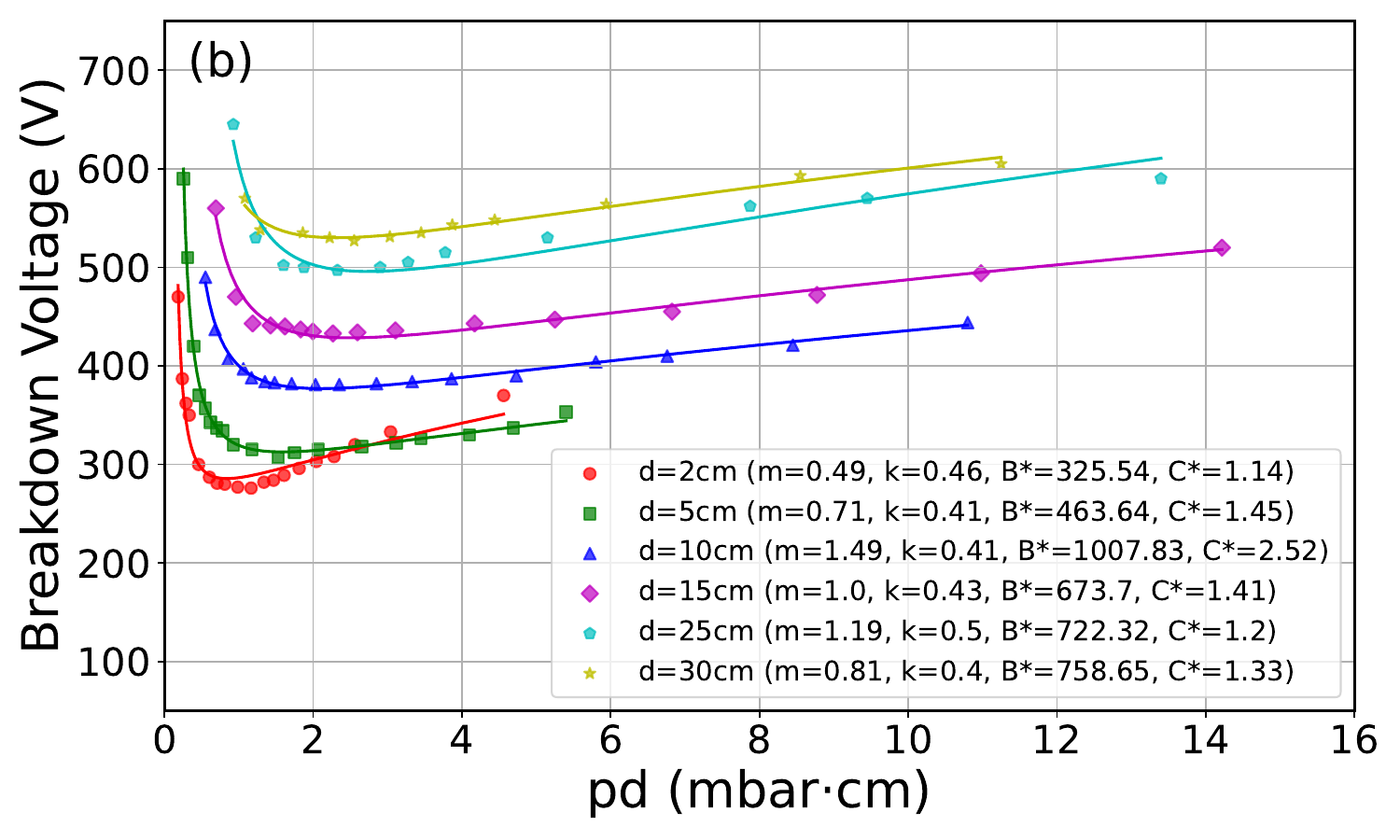}}}
\caption{\label{fig:Fig8} Fitted Paschen curves for (a) symmetric (20 mm cathode and 20 mm anode diameter) configuration and (b) asymmetric (20 mm cathode and 10 mm anode diameter) configuration at different inter-electrode distances.}
\end{figure*}

\begin{figure*} 
 \centering
\subfloat{{\includegraphics[scale=0.4200050]{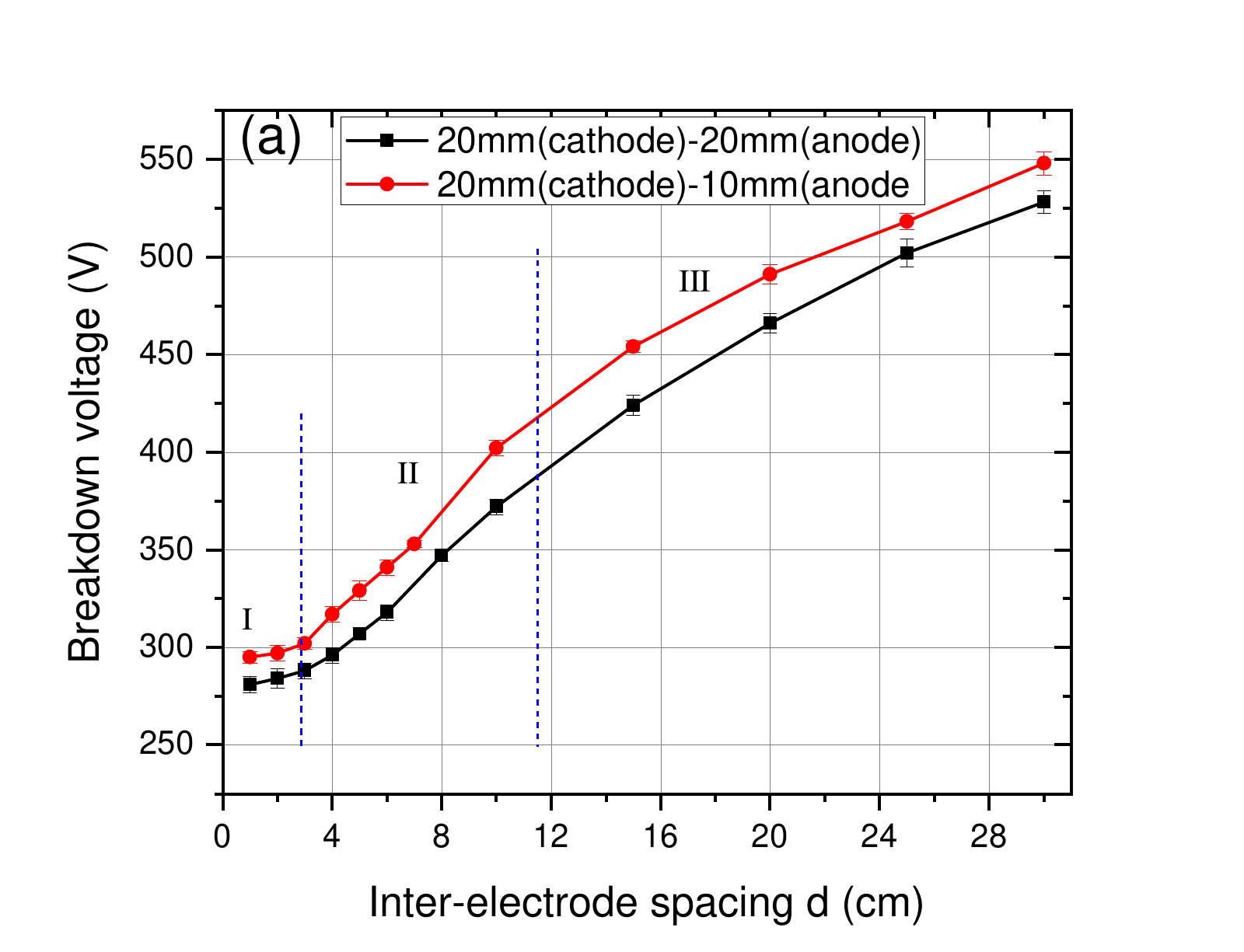}}}%
 \quad
 \subfloat{{\includegraphics[scale=0.420050]{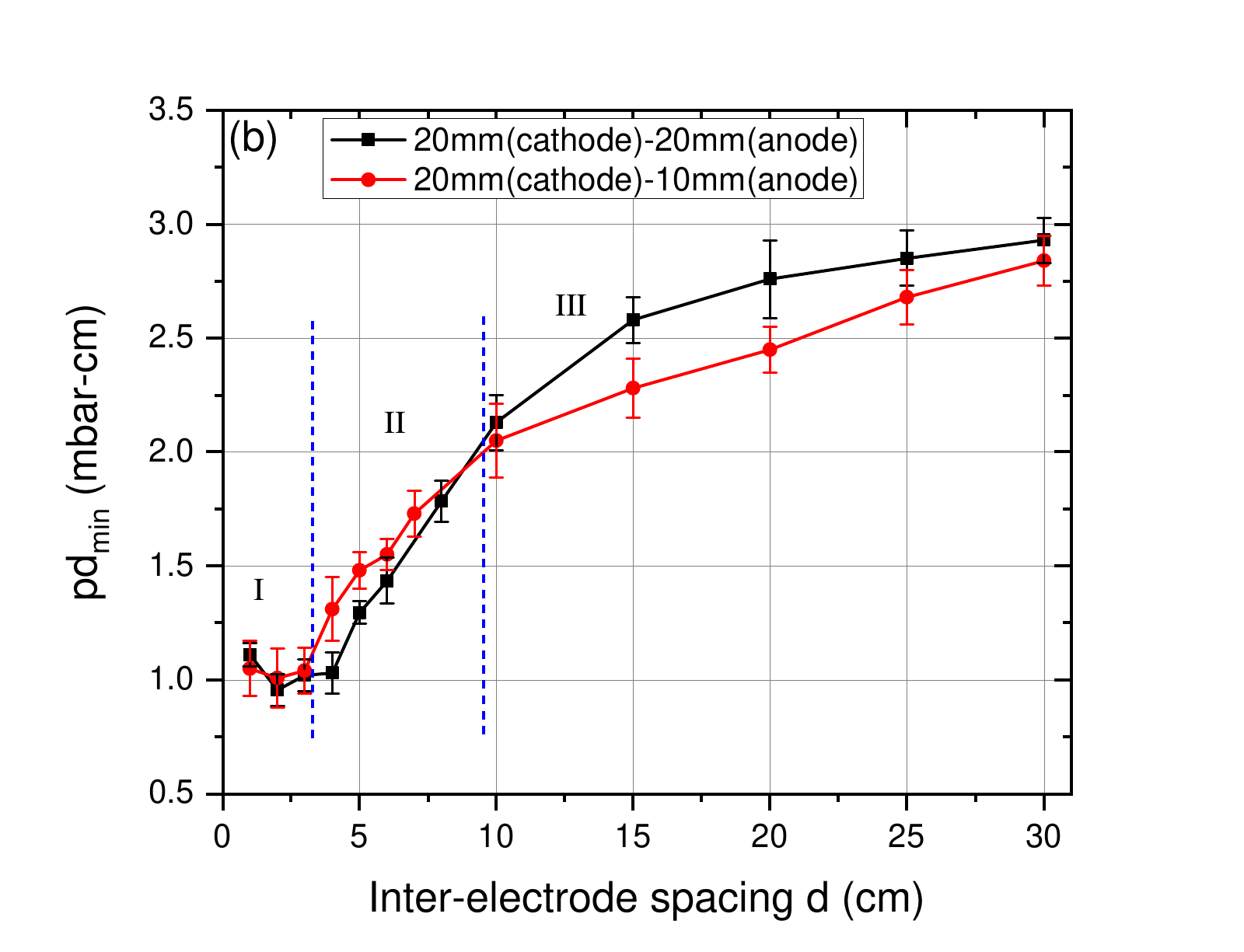}}}
\caption{\label{fig:Fig9}(a) Variation of $V_{B}$, corresponding to minimum value of $pd$ with inter-electrode spacing for symmetric (20 mm cathode and 20 mm anode diameter) and asymmetric electrode (20 mm cathode and 10 mm anode diameter). (b) variation of $pd_{min}$ with inter-electrode spacing for symmetrical and asymmetric electrode configurations.}
\end{figure*}
To explore the role of a smaller-sized cathode on the breakdown of gases, another sets of experiments are performed with a cathode of 20 mm in diameter and different-sized anodes (20 mm and 10 mm in diameter). The Paschen breakdown curves ($V_B$ versus $pd$) for the symmetric and asymmetric electrode configurations are plotted in Fig.\ref{fig:Fig8}(a) and Fig.\ref{fig:Fig8}(b), respectively. The obtained $pd$ curves in either case are fitted with the proposed modified Paschen law. The values of fitted coefficients are slightly different from those recorded in the case of an 80 mm and 40 mm-diameter cathode.\\
The plotted $pd$ curves at different $d$ are used to obtain the minimum breakdown voltage and the minimum value of $pd$ ($pd_{min}$). In the case of a symmetric electrode configuration, $V_{B}$ remains nearly constant up to $d \leq$ 3 cm and then varies linearly with further increasing the inter-electrode spacing. It is clear from Fig.\ref{fig:Fig9}(a) that $V_{B}$ increases at a higher rate when the electrode spacing is between 3 and 11 cm. Then $V_{B}$ increases continuously at a lower rate until the inter-electrode distance reaches 30 cm. The gas breakdown occurs at a comparatively low applied DC voltage ($\sim$280 V) in the symmetric configuration, where both electrodes are close to each other ($d <$ 2 cm). But $\sim$ 240 additional volts are required to achieve the gas breakdown when electrodes are placed 30 cm apart from each other.\\ 
It is understandable from Fig.\ref{fig:Fig9}(a) that $V_{B}$ variation against $d$ for asymmetric electrode configuration follows nearly the same trend that is recorded for the symmetric configuration. In the asymmetric case, gas breakdown takes place at $\sim$300 V when both electrodes are very close to each other ($d <$ 3 cm). But we need a $\sim$550 V voltage difference between cathode and anode if the inter-electrode spacing is about 30 cm. The estimated voltage difference in this case is $\sim$250 V if the inter-electrode distance is increased from 1 cm to 30 cm. A comparative analysis indicates that gas breakdown normally occurs at a lower discharge voltage if a symmetric electrode configuration (20 mm cathode - 20 mm anode) is used in the experiments. A slightly high voltage (20 to 30 V) is required to initiate the gas breakdown phenomena in the asymmetric electrode configuration. It should also be noted that, in either symmetric or asymmetric electrode configurations, the required discharge voltage to initiate the gas breakdown process increases with increasing inter-electrode spacing. It is also noticed from Fig.\ref{fig:Fig9}(a) that there is no significant role of electrode asymmetry on the gas breakdown processes for smaller-sized electrodes. \\
Fig.\ref{fig:Fig9}(b) shows the variation of $pd_{min}$ against $d$ for both electrode configurations. A shifting of $pd_{min}$ from a slightly larger value to a minimum value is observed with changing inter-electrode spacing from 1 to 2 $cm$. After that ($d > 2$ cm), $pd_{min}$ increases at a comparatively higher rate as $d$ increases to 10 cm. When the inter-electrode space is greater than 10 cm, a continuous increase in the $pd_{min}$ value is reported, with a lower rate for both configurations. However, we do not record a significant difference in $pd_{min}$ variation with $d$ for symmetric and asymmetric electrode configurations.
\subsection{Comparative study of Paschen minimum for symmetric configurations}

\begin{figure*} 
 \centering
\subfloat{{\includegraphics[scale=0.4200050]{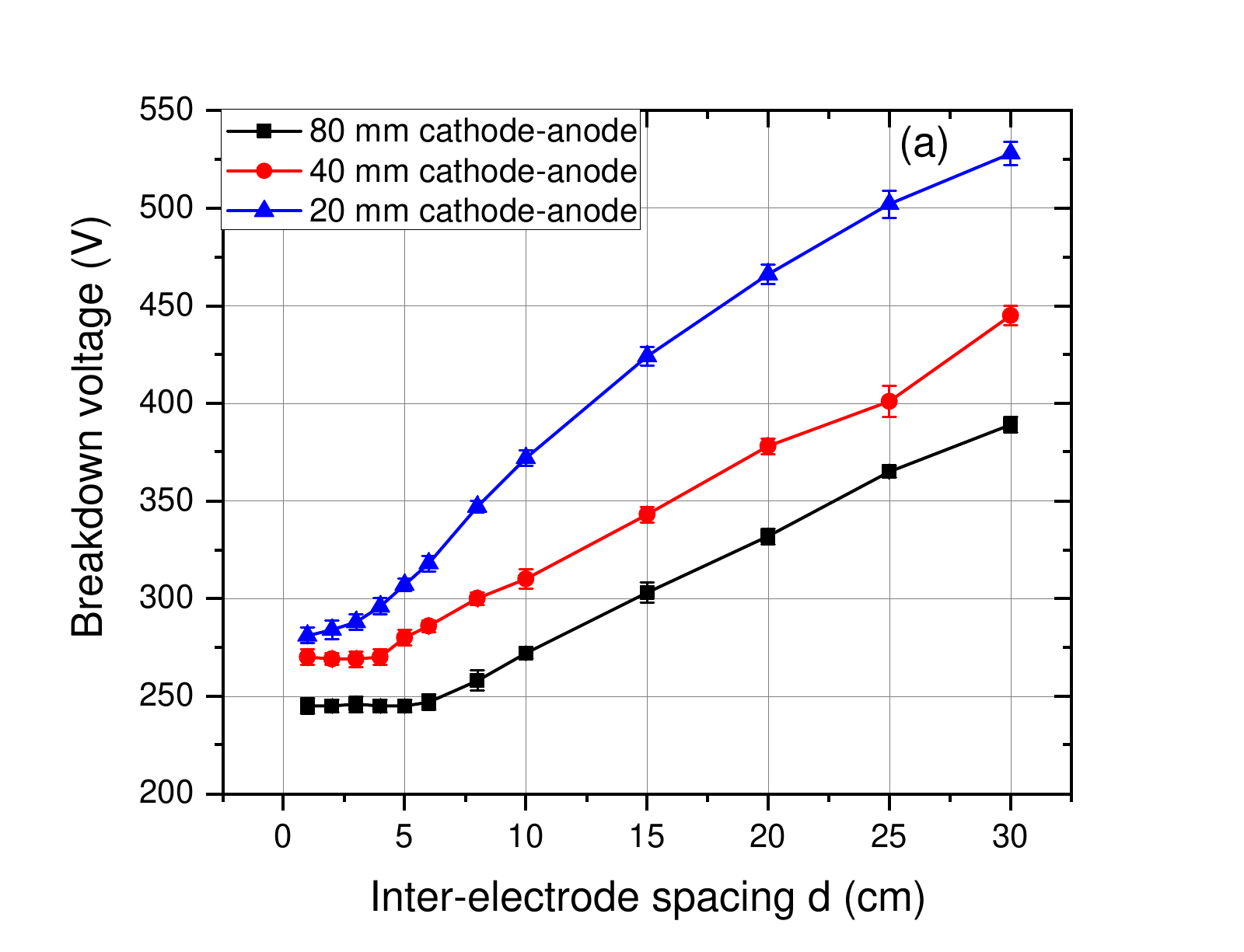}}}%
 \quad
 \subfloat{{\includegraphics[scale=0.420050]{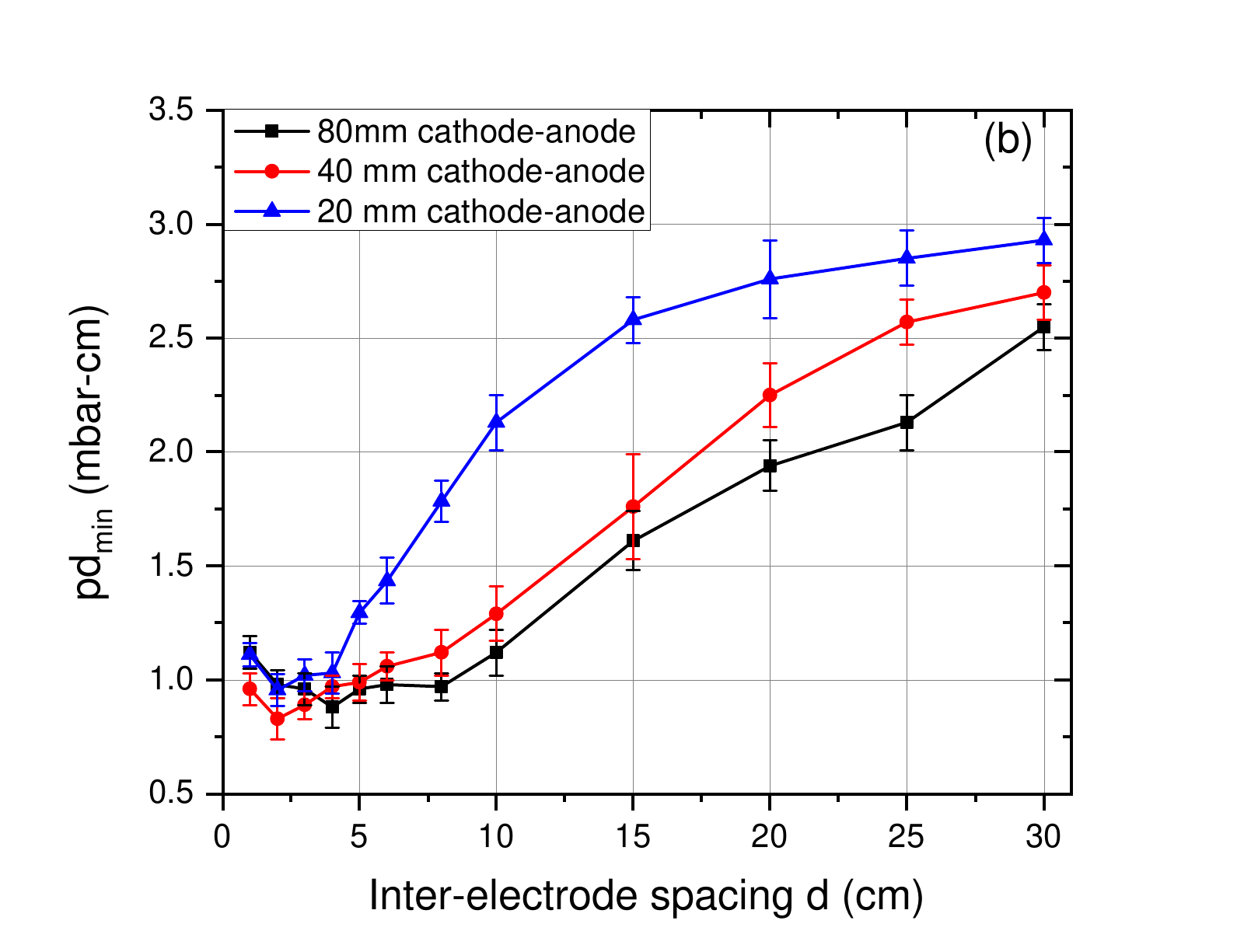}}}
\caption{\label{fig:Fig10}(a) Variation of $V_{B}$, corresponding to the minimum value of $pd$ with inter-electrode spacing (b) variation of pd minimum ($pd_{min}$) with inter-electrode spacing for different symmetric configurations (80 mm - 80 mm, 40 mm - 40 mm and 20 mm - 20 mm).}
\end{figure*}

The experimentally obtained value of $V_{B}$ and $pd_{min}$ for different-sized symmetric electrode configurations are compared in Fig.\ref{fig:Fig10}(a) and Fig.\ref{fig:Fig10}(b), respectively. The gas breakdown occurs at comparatively low discharge voltage in the case of larger-sized electrodes (80 mm in diameter) as compared to smaller-sized electrodes (20 mm in diameter). For a fixed gap of about 5 cm, the VB for 20 mm electrodes is roughly 50 V higher than for 80 mm electrodes, and this difference grows to over 150 V as the inter-electrode spacing is increased to around 30 cm. At small gaps ($d$ = 1–2 cm), the $V_{B}$ difference is about 30 V between 20 mm and 80 mm electrodes, with 40 mm electrodes yielding intermediate values. For large gaps ($d$ = 30 cm), the difference in breakdown voltage between small and large electrodes can reach approximately 150 V. \\
At any inter-electrode distance, the gas breakdown voltage ($V_{B}$) is highest for the 20 mm cathode configuration and decreases with increasing electrode size, causing the Paschen curve to shift vertically with changes in electrode size. The smallest electrodes require higher voltages for breakdown across all gap distances.\\
For $pd_{min}$, the variation is subtler at small gaps (1–5 cm), where it is nearly the same for all configurations. However, as the gap increases, $pd_{min}$ rises faster for the 20 mm electrodes compared to the 40 mm and 80 mm electrodes, indicating that both the minimum breakdown conditions and the position and shape of the Paschen curve strongly depend on electrode size and spacing.
\section{Discussion} \label{sec:sec4}
\begin{figure*}
 \centering
\subfloat{{\includegraphics[scale=0.25]{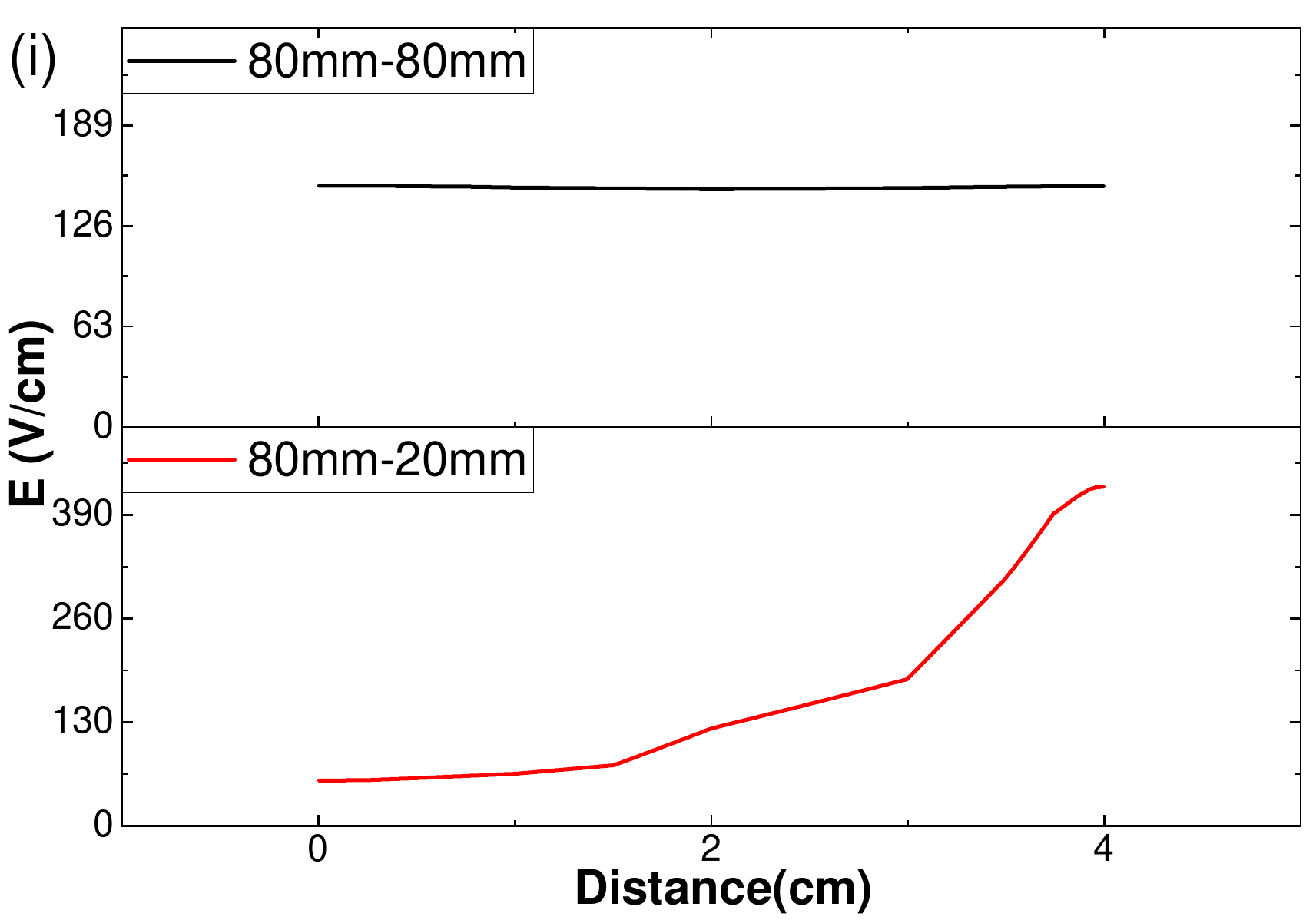}}}%
 \subfloat{{\includegraphics[scale=0.25]{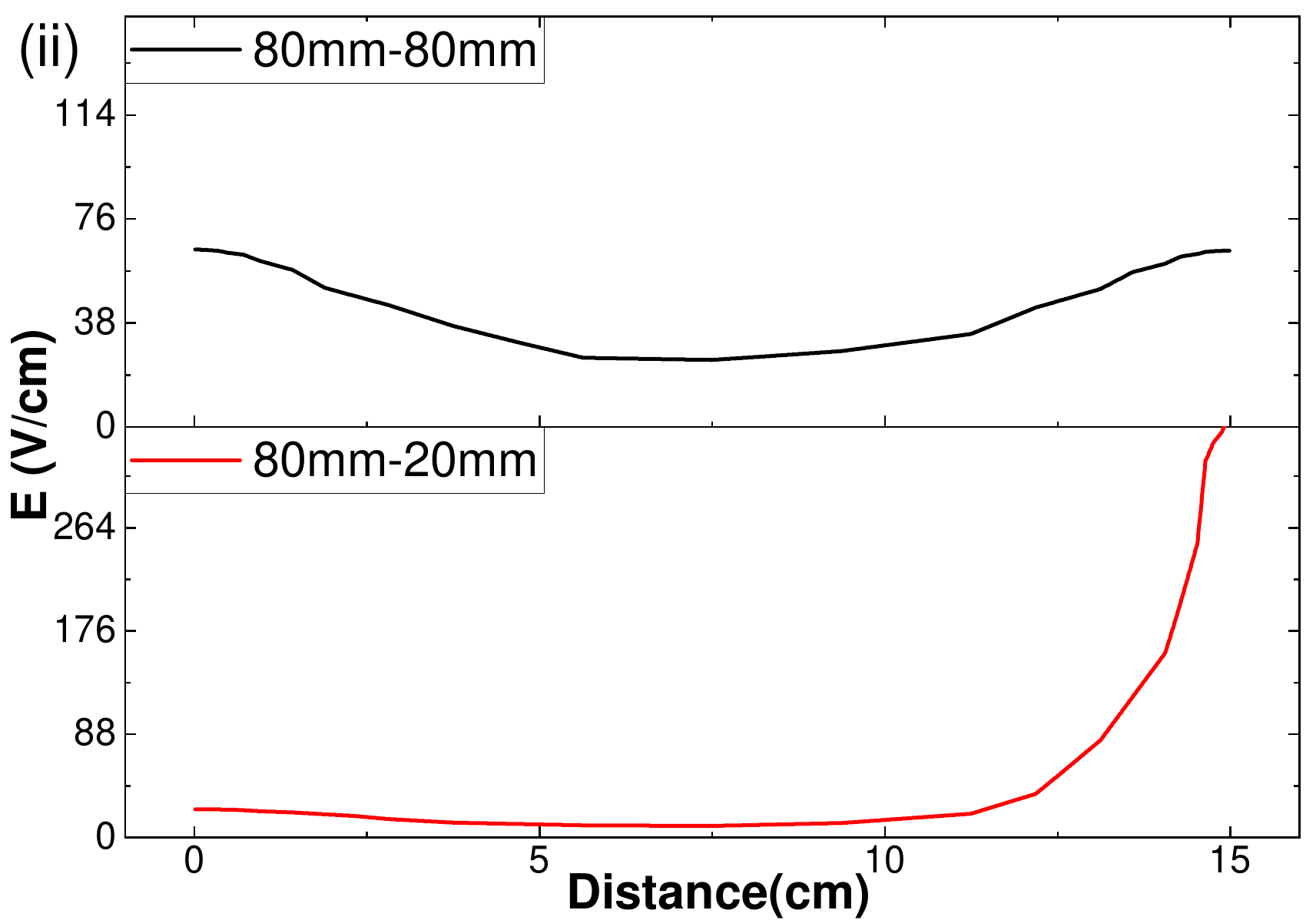}}}
 \quad
 \subfloat{{\includegraphics[scale=0.25]{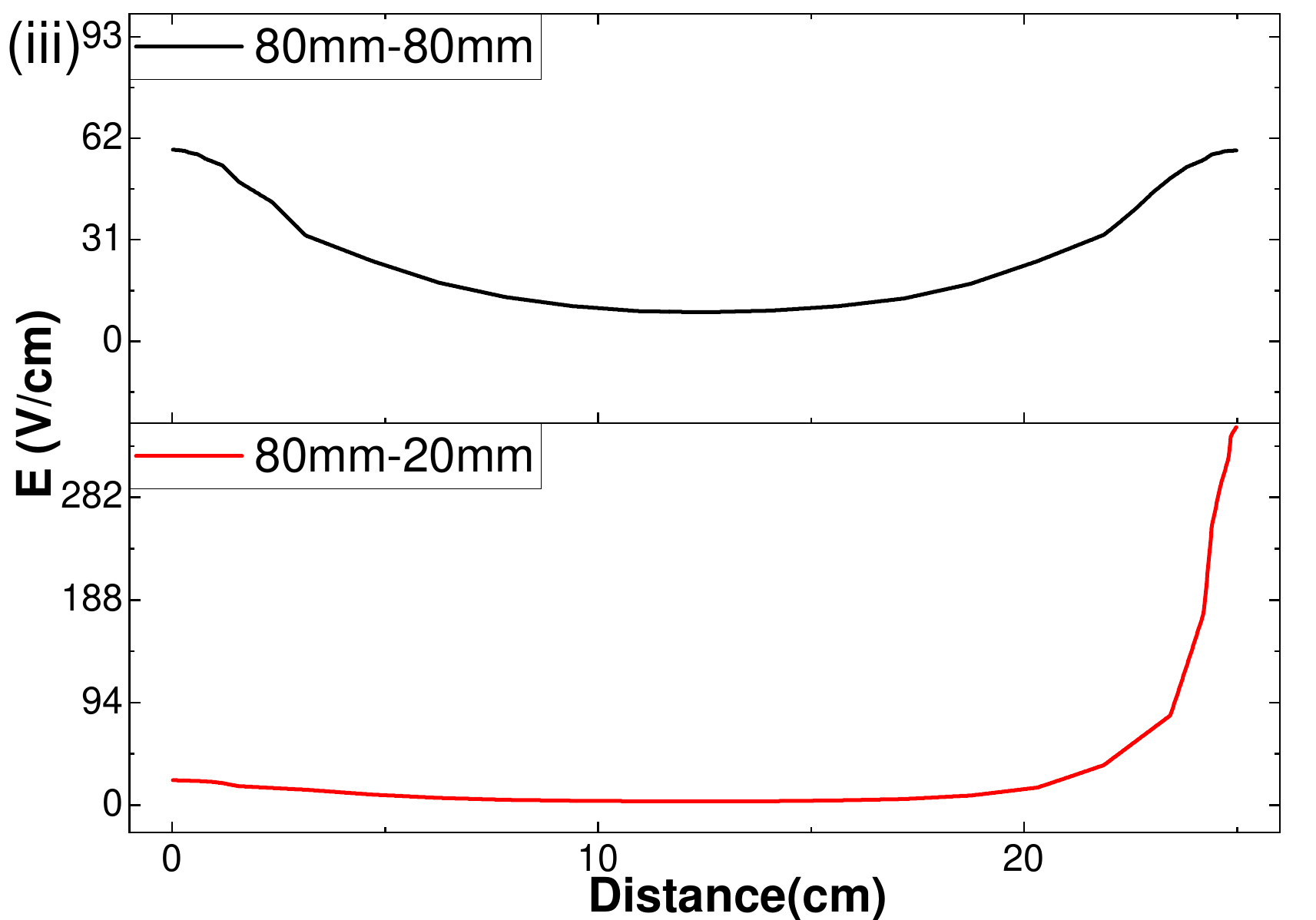}}}
\caption{\label{fig:Fig11} Electric field distribution between the electrodes for (i) $d$ = 4 cm , (ii) $d$ = 15 cm and (iii) $d$ = 25 cm in 80 mm-80 mm configuration and 80 mm - 20 mm configuration.}
\end{figure*}
\begin{figure*} 
 \centering
\subfloat{{\includegraphics[scale=0.25]{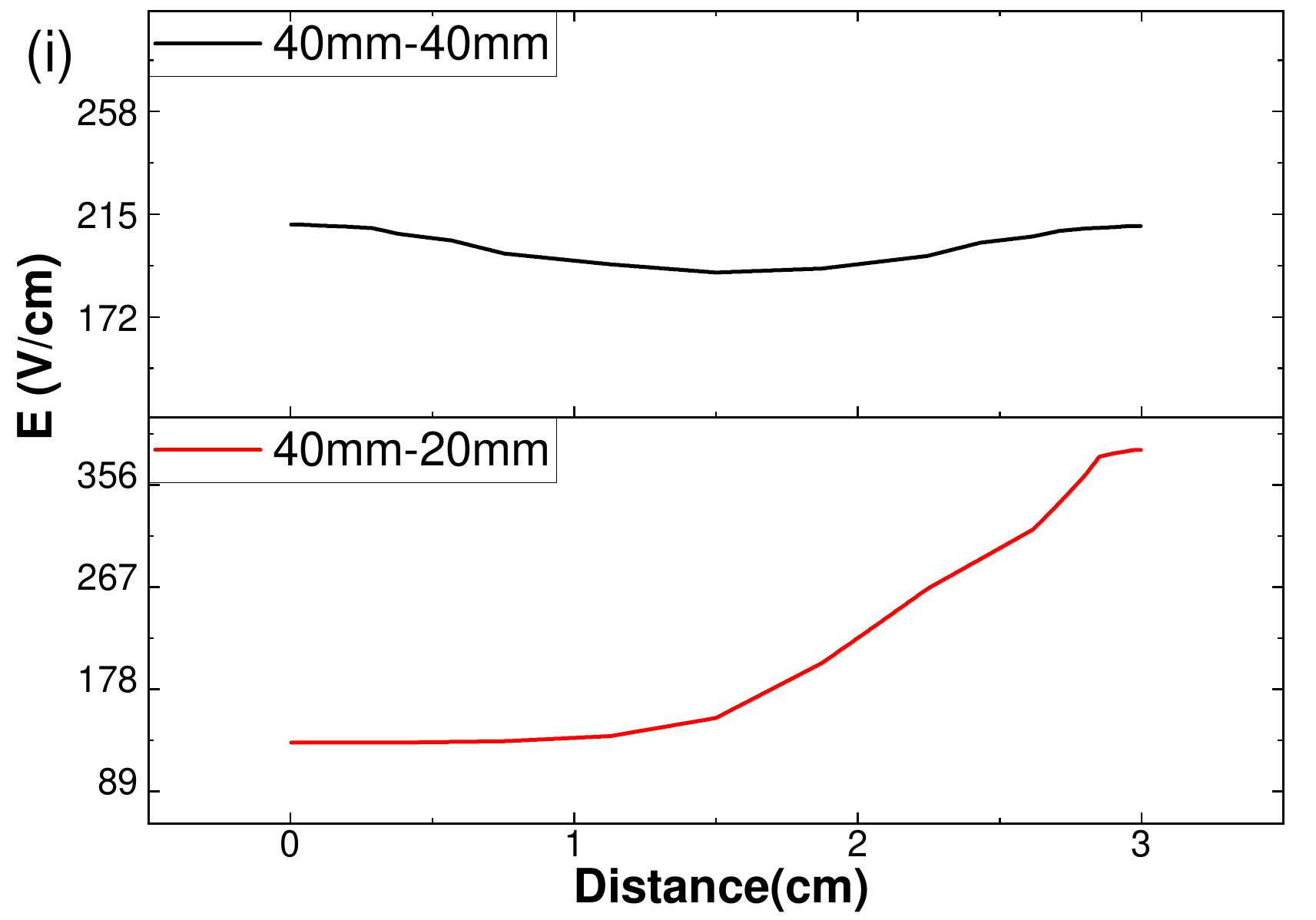}}}%
 \subfloat{{\includegraphics[scale=0.25]{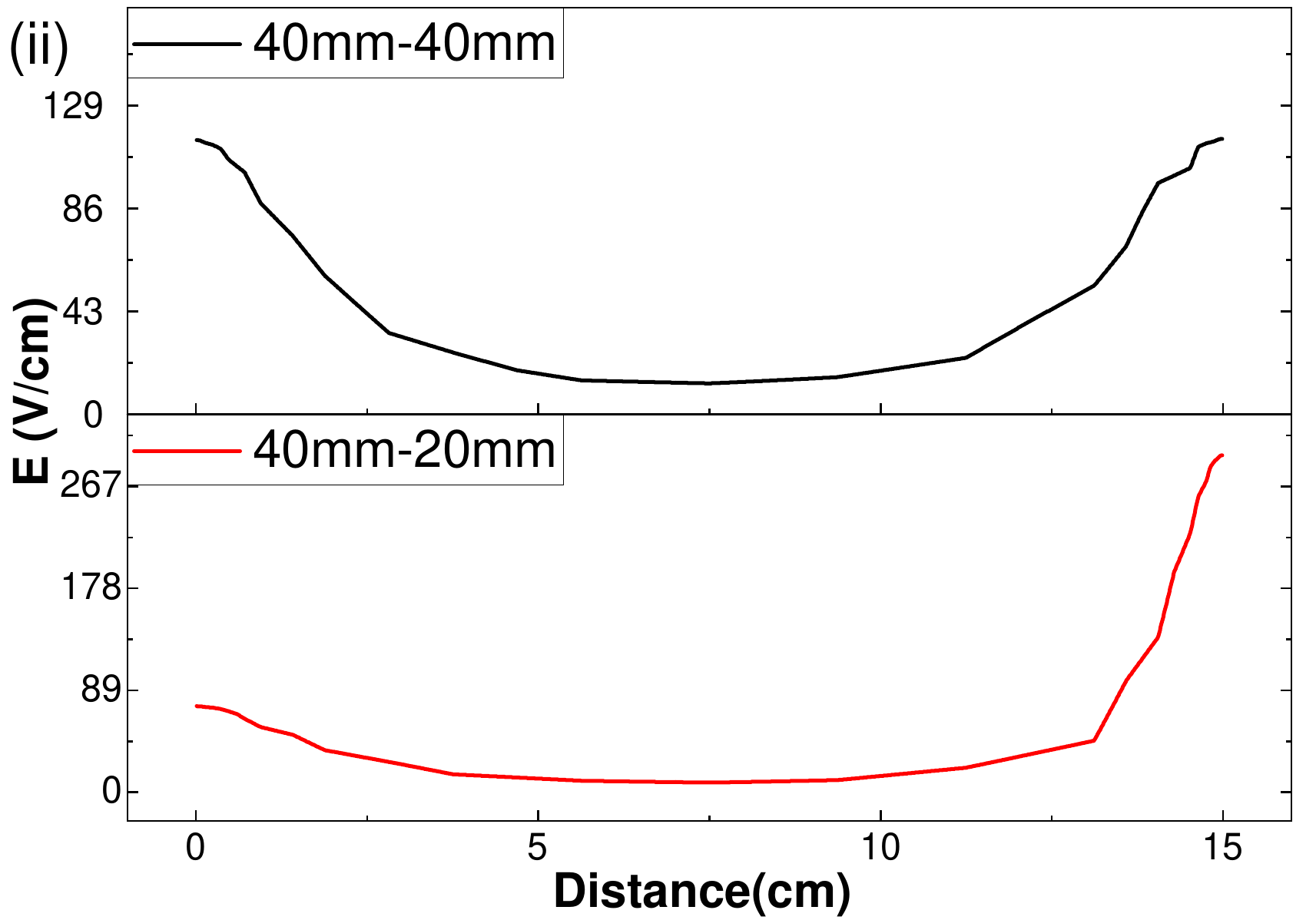}}}
 \quad
 \subfloat{{\includegraphics[scale=0.25]{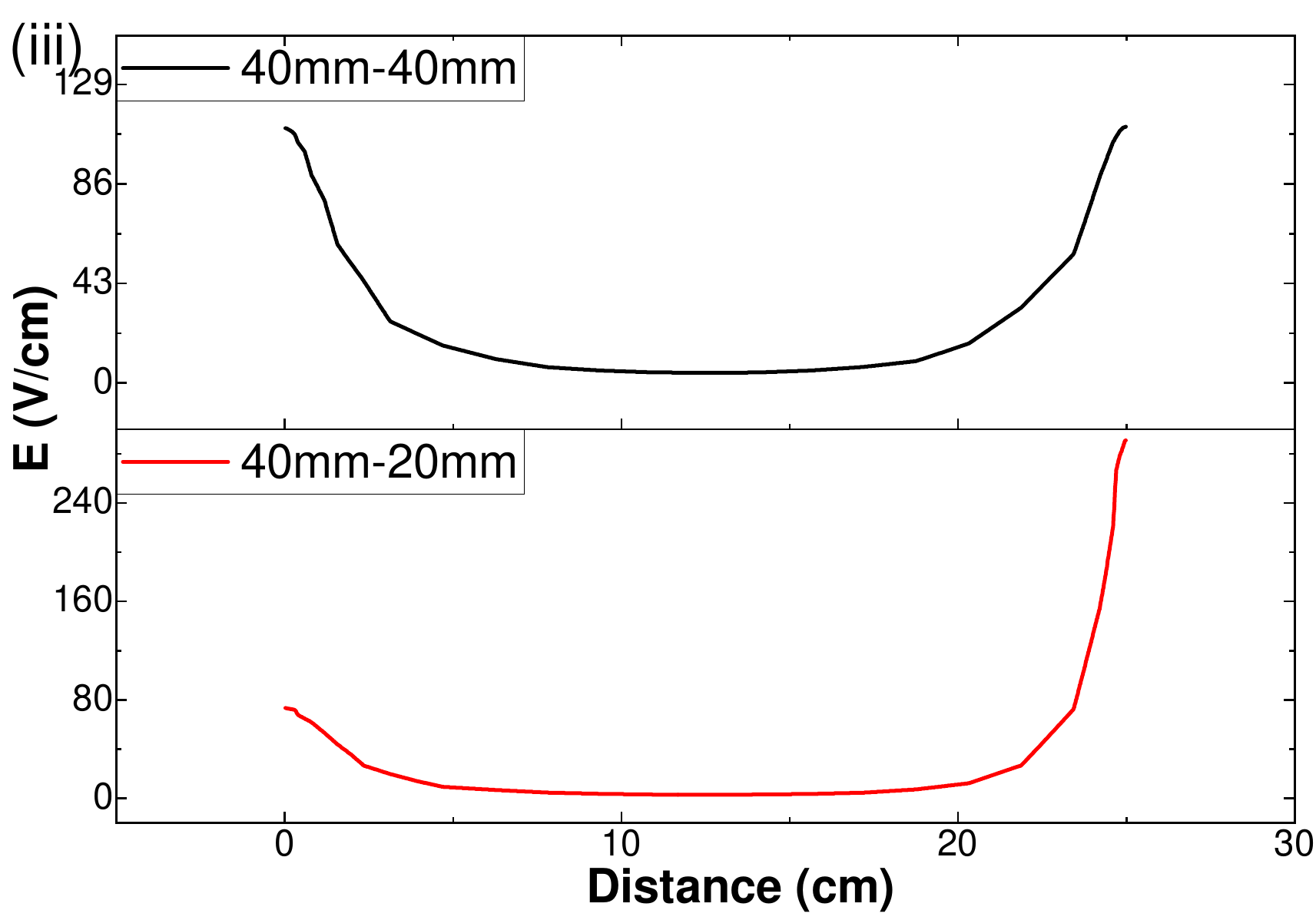}}}
\caption{\label{fig:Fig12} Electric field distribution between the electrodes for (i) $d$ = 3 cm , (ii) $d$ = 15 cm and (iii) $d$ = 25 cm in 40 mm - 40 mm configuration and 40 mm - 20 mm configuration. }
\end{figure*}
\begin{figure*} 
 \centering
\subfloat{{\includegraphics[scale=0.25]{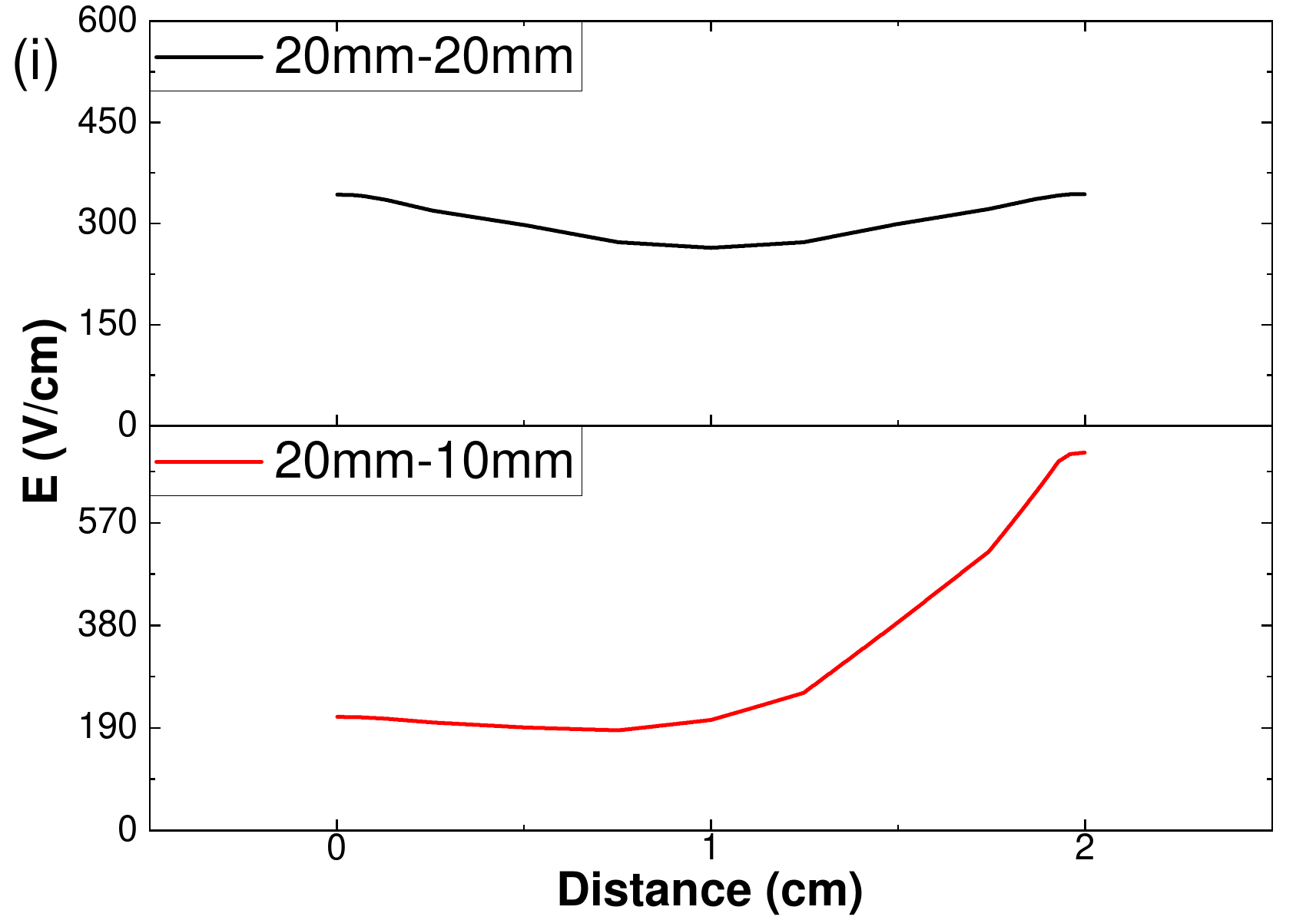}}}%
 \enspace
 \subfloat{{\includegraphics[scale=0.25]{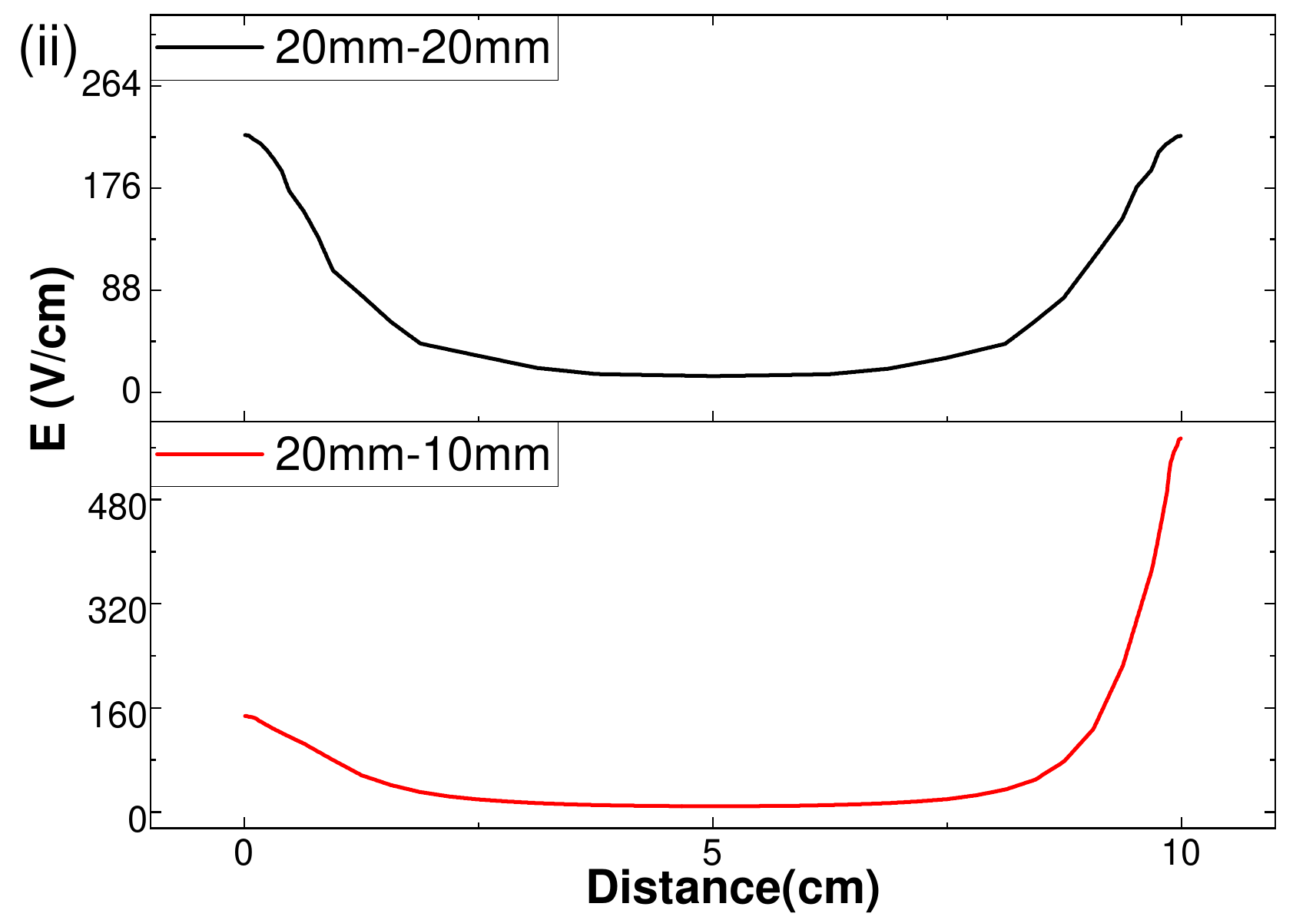}}}
  \quad
 \enspace
 \subfloat{{\includegraphics[scale=0.25]{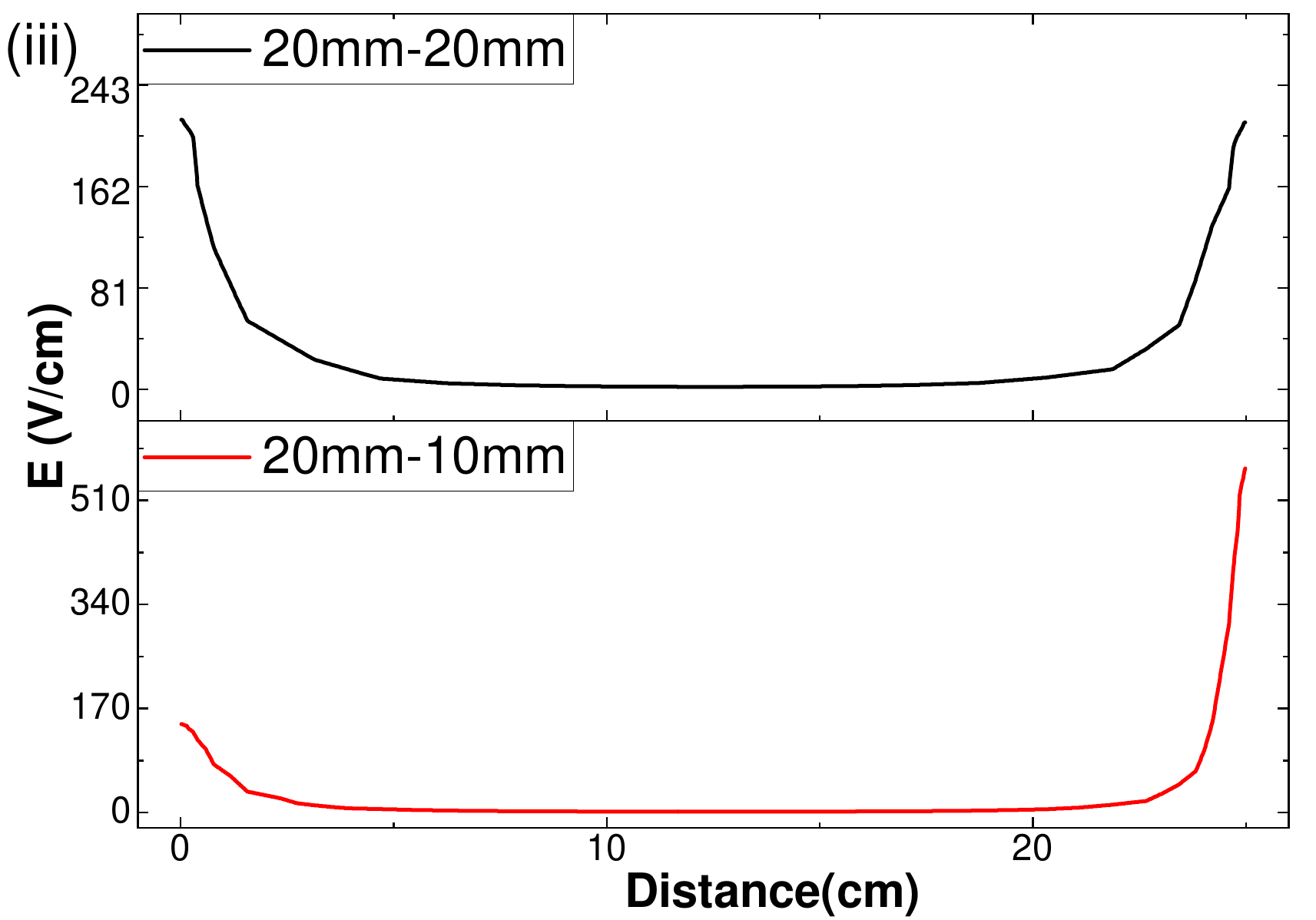}}}
\caption{\label{fig:Fig13} Electric field distribution between the electrodes for (i) $d$ = 2 cm, (ii) $d$ = 10 cm and (iii) $d$ = 25 cm in 20 mm - 20 mm configuration and 20 mm - 10 mm configuration. }
\end{figure*}

It is evident from experimental results, which are presented in Sec.\ref{sec:sec3}, that the Paschen curve is not just a function of $pd$ alone but also depends on the inter-electrode distance, size of electrodes, and configuration of electrodes (symmetric or asymmetric). There are other factors, such as the geometry of the electrodes, the type of electrodes (solid or mesh), and electrode materials, that also affect the gas breakdown processes. The ideal Paschen curve is derived by considering a uniform electric field between electrodes (cathode and anode). Therefore, the electric field distribution (uniform or non-uniform) between biased electrodes strongly influences the gas breakdown processes through ionization and diffusion rates. The electric field distribution between electrodes at different spacings and configurations (symmetric and asymmetric) may be helpful in qualitatively understanding the observed results. By keeping this in mind,  Finite Element Method Magnetics (FEMM) software (\url{https://www.femm.info}) is used to obtain the electric field distribution (2-dimensional) between the cathode and the anode for a fixed applied voltage of 500 V in all cases. The E-field variation between two points on cathode and anode (along axis of cylindrical tube) is plotted for different electrode configurations in Fig.\ref{fig:Fig11}, \ref{fig:Fig12} and \ref{fig:Fig13}.\\
Fig.\ref{fig:Fig11} shows the variation in electric field for symmetric (80 mm - 80 mm) and asymmetric (80 mm - 20 mm) electrode configurations along the electrode gap for  $d=$ 4, 15 and 25 cm. For a symmetric electrode configuration, the electric field remains nearly uniform up to an inter-electrode distance of $\sim$ 5 to 6 cm [Fig.\ref{fig:Fig11}(i)]. The nearly uniform E-field across the gap supports a more efficient ionization process that leads to relatively lower and a constant breakdown voltages. This corresponds to the nearly constant breakdown voltage observed experimentally [Fig.\ref{fig:Fig5}(a)]. On further increasing the inter-electrode distance (15 cm and 25 cm), the electric field in the middle region reduces, causing a gradient with a strong field near the electrodes and a weak field in the central region [see Fig.\ref{fig:Fig11}(ii) and (iii)]. This increase in the weak-E-field region leads to a reduction in ionization and thereby the electron avalanche development for higher inter-electrode distances, causing the breakdown to happen at higher voltage \citep{LISOVSKIYlongtubes}. Therefore, we see a nearly linear increase in the breakdown voltage, as shown in Fig.\ref{fig:Fig5}(a), with increasing the inter-electrode spacing beyond 6 cm.\\
In the asymmetric configuration (80 mm cathode and 20 mm anode), non-uniformity in E-field starts at even smaller electrode separations ($d$ < 5 cm). The electric field becomes stronger and more concentrated near the anode while remaining weaker in the central region and cathode side at a given inter-electrode separation [see Fig.\ref{fig:Fig11}(i)]. This strong E-field near the smaller anode or weak E-field in the central region reduces the effective ionisation rate compared to the symmetric case; consequently, a higher breakdown voltage is required to start a gas avalanche for a given electrode gap, as is depicted in Fig.\ref{fig:Fig5}. The non-uniformity in the E-field, along with some other factors, is responsible for a linear increase in the breakdown voltage when the electrode gap is less than 6 cm. Furthermore, as the inter-electrode distance increases, the E-field gradient becomes more pronounced, maintaining the strongest field near the anode but expanding the weak E-field zone between electrodes even further. The weaker E-field in the central region of the discharge gap [Fig.\ref{fig:Fig11}(ii)-(iii)] reduces the primary ionization rates, and the asymmetry increases the diffusion of charged particles from the discharge volume (a conical discharge volume is expected in the asymmetric case). Therefore, a gas avalanche is expected at a higher applied voltage at a large inter-electrode spacing [see Fig.\ref{fig:Fig5}] compared to the symmetric case.\\
For a 40 mm - 40 mm symmetric electrode configuration, the electric field between electrodes is nearly uniform for small inter-electrode distances, specifically up to about 3–4 cm [Fig.\ref{fig:Fig12}(i)]. In this region, the breakdown voltage is observed to remain low and fairly constant [Fig.\ref{fig:Fig7}(a)]. As the separation increases beyond 5 cm, the E-field profile begins to change, with strong gradients developing near the electrode surfaces and a weak E-field region expanding in the central region of the discharge gap. For intermediate gap values ($<$ 12 cm), the electric field in the middle of the electrode gap is not as weak as in larger configurations (d = 25 cm), so the breakdown voltage increases at slightly different rates. At greater separations, the central region’s field weakens much more substantially, causing the breakdown voltage to rise more sharply as a larger applied voltage is needed to sustain electron avalanches and initiate gas breakdown [see Fig.\ref{fig:Fig7}(a)]. 
However, for the 40 mm - 20 mm asymmetric electrode configuration, the non-uniformity in the E-field begins below d < 4 cm. Near both electrodes, the E-field strength is non-uniform, with a notably stronger electric field concentrated near the smaller 20 mm anode. This effect reflects in $V_B$ variation below d $<$ 4 cm in Fig.\ref{fig:Fig7}(a). As inter-electrode spacing increases (beyond 4 cm), the field strength becomes weaker in the middle discharge region while remaining strong near the electrodes (strong near the anode). This non-uniformity in the E-field reduces the effective ionization rate and increases the diffusion rate in the bulk discharge region. Hence, high voltage is required to initiate gas breakdown in case of larger inter-electrode spacing [see Fig.\ref{fig:Fig7}(a)]. 
\\
In case of 20 mm - 20 mm symmetric electrode configuration, the E-field is nearly uniform up to $d$ = 2 or 3 cm [see Fig.\ref{fig:Fig13}(i)]. The uniformity in E-field causes a more consistent acceleration of electrons, leading to lower breakdown voltages [see Fig.\ref{fig:Fig9}(a)]. On further increasing the inter-electrode distance (beyond 3 cm), the E-field strength in the middle region of the discharge gap decreases and will reach a minimum value when $ d=$ 30 cm. The length of the weak E-field region increases with the increase in inter-electrode distance [Fig.\ref{fig:Fig13}(ii)-(iii)]. The weak E-field in the bulk discharge gap reduces ionization rates, so more voltage is required to start an avalanche at large inter-electrode distances, as depicted in Fig.\ref{fig:Fig9}(a). \\ 
For the 20 mm - 10 mm asymmetric electrode configuration, the non-uniformity in the E-field starts below $d<$ 2 cm. The strength of the E-field is also non-uniform near both electrodes [Fig.\ref{fig:Fig13}(i)]. This effect reflects in an increasing trend of $V_B$ for smaller inter-electrode distance $d < 3$ cm, as shown in Fig.\ref{fig:Fig9}(a). With increasing distance (beyond 3 cm), the stronger E-field region is concentrated near the electrodes, especially the anode, and the weak E-field is extended in the discharge region. This spatial variation causes fewer ionizing collisions in the weak field regions, increasing the overall breakdown voltage with increasing inter-electrode spacing. For asymmetric and symmetric configurations, the discharge volume is in the form of a cone and a cylinder, respectively. The diffusion rate of plasma particles is expected to be higher in the case of canonical discharge volume compared to cylindrical volume, resulting in higher breakdown voltage for asymmetric electrode configuration at a given d [Fig.\ref{fig:Fig9}(a)]. \\\\
In the case of symmetric configurations with different electrode sizes (80 mm-80 mm, 40 mm-40 mm, and 20 mm-20 mm), the difference in E-field strength and volume of discharge regions determines the gas breakdown voltage at a given inter-electrode spacing. The higher expected diffusion rates of plasma particles (before Townsend avalanches) and comparatively lower ionization rates due to a weaker E-field in the bulk discharge gap in smaller-sized electrode configurations (20 mm -20 mm) favour higher breakdown voltage compared to 80 mm-80 mm symmetric configurations [see Fig.\ref{fig:Fig10}].   
It should be noted that the results of the present work are qualitatively understood based on the electric field distribution for different cases. However, other factors such as anode material \citep{adams2017effect}, roughness of electrode surface \citep{adams2017effect}, orientation of electrodes, composition of gases, diffusion of charged particles etc., also affect the gas breakdown processes and possible sources to modify pd curves.    

The theoretical value of $pd_{min}$ obtained from Eq.\ref{paschen law OG} by finding out the turning point of the curve:
\begin{equation}
    pd_{min} = \frac{e\ln{(1+\frac{1}{\gamma_{se}}})}{A}
\end{equation}
As per the classical Paschen law, $A$ and $\gamma_{se}$ are constant, so $pd_{min}$ should also be a constant. However, our experimental results show that these coefficients vary throughout the measurements. Also, the coefficient $A$ depends upon the ionization cross section. For larger inter-electrode gaps, the electric field in the central region will be weak, reducing the electron energy and making it insufficient for effective ionization. Therefore, the ionization cross-section decreases, and this in turn contributes to the increase in the $pd_{min}$ at larger inter-electrode distances as observed in the present study [Fig.\ref{fig:Fig10}(b)]. For a very small inter-electrode distance, the E field will be too high, causing the ionization cross-section to be large owing to the lower $pd_{min}$ observed. In the symmetric configuration [see Fig.\ref{fig:Fig10}(b)], the 20 mm configuration has a higher $pd_{min}$ because, as the electrode size reduces, the electric field in between the electrodes decreases, causing the $pd_{min}$ to increase. It is still not clear how the $pd_{min}$ varies at smaller distances when the inter-electrode distance is comparable with the radius of the electrodes. 
\section{Summary and future outlook} \label{sec:sec5}
In this study, we investigated the breakdown of argon gas at different inter-electrode distances for symmetric and asymmetric electrode size configurations. The key findings of these investigations are as follows-
\begin{itemize}
    \item Classical Paschen law is found to be inadequate in describing the breakdown characteristics of gas for asymmetric and symmetric electrode configurations if inter-electrode spacing is greater than the size of the cathode. However, it describes the breakdown behaviour of gas for larger-sized symmetric electrodes (e.g., 80 mm - 80 mm) when $d <$ 80 mm.
    \item A modified empirical relation incorporating power-law dependencies, which better captures deviations from the classical law, is proposed and fitted to experimental data. A strong dependency of the fitting parameters on electrode configurations and inter-electrode spacing is observed.
    \item The gas breakdown happens at a higher applied voltage in asymmetric electrode configurations compared to the symmetric-sized electrode configuration.  
    \item The gas breakdown strongly depends on the size of the electrodes, especially the size of the cathode. The gas breakdown voltage increases with a decrease in electrode size. 
    \item The breakdown (corresponds to the minimum $pd$ value) voltage increases approximately linearly for asymmetric electrode configurations as the inter-electrode gap increases. In symmetric configurations, it starts to increase linearly once the inter-electrode gap exceeds the cathode size.
    \item The minimum $pd$ ($pd_{min}$) value shifts to a lower value and then starts to attain the original value, and then increases as the inter-electrode distance changes from minimum (1 cm) to maximum (30 cm) for both electrode configurations. 
\end{itemize}
The experimentally observed results are supported by a finite element analysis of the electric field distribution between the cathode and anode for a given electrode configuration. The non-uniformity in the E-field, whether in a symmetric or asymmetric electrode configuration in the discharge gap, qualitatively explains the shifting of the Paschen minimum and increase in breakdown voltages with increasing the length of the discharge gap. It is expected that the breakdown voltage depends on the volume of the discharge zone (available volume for ionization), which could be an additional factor, along with the E-field, in the bulk discharge regime for a difference in breakdown voltage for larger and smaller-sized electrodes. The diffusion of charged particles from the discharge zone (discharge gap) also affects the gas breakdown process in both symmetric and asymmetric electrode configurations, which is not within the scope of the present study. For a better understanding of observed results, detailed computational simulations of the present setup are needed, which could be a future project. 
 \section{Acknowledgement} The authors sincerely thank Mr Sumit Tomar for his technical help in developing the plasma setup in the plasma physics laboratory. Dr Choudhary thanks Dr Neeraj Chaubey and Dr Vikram Singh Dharodi for their valuable support in developing the VMC DC discharge device. The authors also extend their sincere thanks to the physics workshop and store personnel for their constant support during the development of this plasma device at the Department of Physics and Astrophysics, University of Delhi.     
\section{Author Declarations}
\subsection{Conflict of Interest}
The authors declare no conflicts of interest.
\subsection{Author Contributions}
Dr Choudhary conceived the idea for this project and performed the experiments. Ms Hridya performed the literature search to fit and explain the experimental data. The first draft of the manuscript was written and revised by both authors. Both authors read and approved the final manuscript.   
\section{Data Availability}
The datasets generated during and/or analyzed during the current study are available from the corresponding author upon reasonable request.
\bibliographystyle{jpp}
\bibliography{References}
\end{document}